\documentclass[11pt]{article}

\usepackage{epsfig}
\usepackage{pst-grad} 
\usepackage[space]{grffile} 
\usepackage{etoolbox} 
\makeatletter 
\patchcmd\Gread@eps{\@inputcheck#1 }{\@inputcheck"#1"\relax}{}{}
\makeatother

\usepackage{graphicx}
\usepackage{amsmath}
\usepackage{psfrag}
\usepackage{color,soul}
\usepackage{natbib}
\usepackage{float}
\usepackage{overpic}
\usepackage{tikz}
\usepackage{multirow}
\usepackage{bm}
\usepackage{textgreek}
\usepackage{url}
\usepackage{amssymb}
\usepackage{xfrac}
\usepackage{setspace}
\usepackage{gensymb}
\usepackage{mathtools}
\usepackage{arydshln}
\usepackage{MnSymbol} 

\usepackage{booktabs} 
\usepackage{algorithm}
\usepackage[noend]{algpseudocode}

\def \eg{\emph{e.g.}, }

\def \Dt{$\Delta t$~}

\providecommand{\Rey}{\textit{Re}}
\providecommand{\umu}{\ensuremath{\mu}}

\usepackage[noblocks]{authblk}
\usepackage[margin=1in]{geometry}
\usepackage[title]{appendix}

\newcommand\affiliation[1]{\gdef\@affiliation{\let\aff\aff@inst#1}}
\gdef\@affiliation{}
\def\email#1{Email address for correspondence: #1}
\def\aff#1{\ignorespaces\textsuperscript{#1}}
\def\corresp#1{\unskip\thanks{#1}}
\numberwithin{equation}{section}

\renewenvironment{abstract}
{\begin{quote}
\noindent \rule{\linewidth}{.5pt}\par{\bfseries \abstractname.}}
{\medskip\noindent \rule{\linewidth}{.5pt}
\end{quote}
}

  \DeclareTextFontCommand\textsfi{\usefont{OT1}{cmss}{m}{sl}}
  \DeclareMathAlphabet\mathsfi            {OT1}{cmss}{m}{sl}
  \DeclareTextFontCommand\textsfb{\usefont{OT1}{cmss}{bx}{n}}
  \DeclareMathAlphabet\mathsfb            {OT1}{cmss}{bx}{n}
  \DeclareTextFontCommand\textsfbi{\usefont{OT1}{cmss}{m}{sl}}
  \DeclareMathAlphabet\mathsfbi            {OT1}{cmss}{m}{sl}
  \IfFileExists{t1phv.fd}
  {\DeclareTextFontCommand\textsfbi{\usefont{T1}{phv}{b}{it}}
  \DeclareMathAlphabet\mathsfbi            {T1}{phv}{b}{it}
  }{}
  \IfFileExists{ot1phv.fd}
  {\DeclareTextFontCommand\textsfbi{\usefont{OT1}{phv}{b}{it}}
  \DeclareMathAlphabet\mathsfbi            {OT1}{phv}{b}{it}
  }{}
  
\DeclareSymbolFont{matha}{OML}{txmi}{m}{it}

\usepackage{hyperref}
\hypersetup{
    colorlinks = true,
    urlcolor   = blue,
    citecolor  = black,
}

\usepackage{hyperref}

\definecolor{darkblue}{rgb}{0,0,0.80}

\hypersetup{
    colorlinks=true,
    linkcolor={darkblue},
    citecolor={darkblue},
    filecolor=black,
    urlcolor={darkblue},
    plainpages=false,
}

\title{\bf Wave interactions in a screeching jet}

\author[1]{\bf Ali Farghadan}
\author[2]{\bf Jayson Beekman}
\author[2]{\bf Petr\^onio Nogueira}
\author[2]{\bf Daniel Edgington-Mitchell}
\author[1]{\bf Aaron Towne\corresp{\email{towne@umich.edu}}}

\affil[1]{\normalsize Department of Mechanical Engineering, University of Michigan, Ann Arbor, MI, USA }
\affil[2]{\normalsize Department of Mechanical and Aerospace Engineering, Monash University, VIC 3800, Australia}

\date{}
\begin{document}
\maketitle

\begin{abstract}
We use a series of global models to investigate the linear and nonlinear interactions between shock cells, Kelvin-Helmholtz waves, guided jet modes, and other fluctuations in a screeching jet.  First, we identify a set of lightly damped global eigenmodes of the Navier-Stokes operator linearized about the mean flow and show that they result from interactions with different shock-cell wavenumbers.  Second, we use resolvent analysis to study the linear input-output behavior of the jet and obtain a time-periodic representation of the screech mode, which compares favorably with experimental data.  Third, we use harmonic resolvent analysis to study triadic interactions, including inter-frequency energy transfer, between the screech mode determined from resolvent analysis and other fluctuations in the jet. The components of the optimal harmonic resolvent mode at harmonics of the screech frequency are consistent with experimental observations that have not been previously predicted by global models.  Fourth, we leverage a novel bilinear formulation of harmonic resolvent analysis to study the impact of the screech mode's nonlinear self-interaction on other fluctuations in the jet.  We show that the forcing provided by this nonlinear self-interaction of the screech mode, along with its triadic interactions with other frequencies embedded within the harmonic resolvent operator, is sufficient to explain the redistribution of energy to other frequencies and the associated experimental observations. In aggregate, these findings underscore the critical role of triadic and nonlinear interactions in shaping screech dynamics and offer a promising workflow for studying similar interactions in other flows dominated by periodic motions.
\end{abstract}


\section{Introduction}

Screeching jets are characterized by intense acoustic emissions \citep{Powell53}, resulting from a feedback loop between downstream-traveling Kelvin-Helmholtz (KH) waves and upstream-traveling acoustic-like waves, known as guided jet modes, coupled by shock cells \citep{nogueira2022absolute}. In aerospace engineering, this phenomenon significantly impacts aircraft operations by contributing to structural fatigue, increasing maintenance requirements, and potentially shortening the operational lifespan of critical components \citep{Tam95}.  Addressing jet screech is essential for designing effective exhaust systems and noise mitigation strategies, especially as demands for noise reduction intensify near urban areas and airports \citep{Basneretal17}. Beyond aviation, screeching jets pose challenges in industrial processes involving high-speed exhaust flows, where noise control is vital for regulatory compliance and environmental protection \citep{TamTanna82, Raman99}.  Due to its multidisciplinary nature, involving acoustics, fluid dynamics, and materials science, jet screech has motivated extensive experimental and computational research aimed at understanding and mitigating these challenges.

The physics behind jet screech are driven by several interacting mechanisms, including shock waves, Kelvin-Helmholtz (KH) instability waves, and acoustic guided jet modes \citep{Edgingtonetal21}. Shear-layer instabilities form downstream-traveling KH wavepackets, which interact with shock cells to generate both upstream- and downstream-traveling waves, including upstream-traveling guided jet modes \citep{nogueira2024guided}.  At certain frequencies, the KH waves and guided jet modes resonate, leading to the observed screech tones.  

Existing models of jet screech differ in which wave interactions they explicitly consider. Some approaches describe sound generation as arising from interactions between KH waves and the quasi-periodic shock-cell structure without directly considering an upstream wave to close the resonance loop, an approach originally developed in the context of broadband shock-associated noise and later interpreted as providing the fundamental mechanism underlying screech in limiting cases \citep{TamTanna82,tam1987stochastic}. Other models describe screech as a closed resonance loop bounded by an upstream reflection point at the nozzle lip and a downstream reflection point near the third or fourth shock cell, with the screech frequency selected by a phase-matching condition over the loop length \citep{powell1953mechanism, tam1986proposed, Mancinellietal19}. More recent studies have shown that the upstream-travelling component responsible for closure is better described as a guided jet mode rather than a free-field acoustic wave \citep{Gojonetal18, Edgingtonetal18}, leading to revised resonance models with improved agreement with observed screech frequencies and staging behavior \citep{Mancinellietal19, mancinelli2021complex, Edgingtonetal22}. Within this framework, the generation of the upstream-travelling wave is often attributed to scattering or triadic interactions between the KH wavepacket and the shock-cell structure, but the shock cells are typically not explicitly included in the analysis. Finally, global stability and absolute-instability analyses treat screech as a self-excited mode of a shock-containing jet, with the shock-cell pattern embedded in the base or mean flow and upstream- and downstream-travelling waves emerging naturally from the eigenstructure, rather than being prescribed through a reduced feedback loop \citep{beneddine2015global, Edgingtonetal21, nogueira2022absolute}.

In this paper, we use a series of linear and nonlinear models to investigate the interactions among different types of structures, or waves, in screeching jets.  First, we use a global linear stability analysis to identify lightly damped discrete eigenmodes of the Navier-Stokes equations linearized about the mean jet.  This approach is well-established in the recent literature \citep{beneddine2015global, Edgingtonetal21}.  Unlike these previous works, we find multiple lightly damped modes, which we show correspond to interactions between the Kelvin-Helmholtz and guided jet modes and multiple peaks in the shock-cell wavenumber spectrum \citep{nogueira2022closure}.  The least damped mode matches the experimentally observed screech frequency.  

Second, we perform a resolvent analysis of the same linear operator to study the linear input-output behavior of the jet and obtain a time-periodic representation of the screech mode suitable for further analysis.  Resolvent analysis provides frequency-dependent forcing and response modes that maximize the gain between them and has been fruitfully used to study many turbulent flows \citep{MckeonSharma10, Towneetal18, jovanovic2021bypass}, including jets \citep{Schmidtetal18, lesshafft2019resolvent}.  \citet{Liang2024active} recently applied resolvent analysis to a screeching planar jet; the current paper, to our knowledge, is the first global resolvent analysis of a screeching circular jet.  We show that the resolvent gain spectrum for the screeching jet contains sharp peaks associated with the discrete screech eigenvalues for the global stability analysis.  The leading forcing and response modes at the dominant screech frequency reveal the structures driving the feedback loop between Kelvin-Helmholtz wavepackets, guided jet modes, and shock-cell interactions.  The highest gain mode occurs at the screech frequency and closely matches the leading proper orthogonal decomposition \citep{Lumley67, Sirovich87_1} mode extracted from particle-image velocimetry (PIV) data.  

Third, we use harmonic resolvent analysis \citep{Padovanetal20} to study the triadic interactions between the screech mode and other fluctuations in the jet.  In harmonic resolvent analysis, the Navier-Stokes equations are linearized about a time-periodic base flow, which in our case consists of the sum of the mean flow and the screech mode obtained from resolvent analysis.  Transforming the resulting linear time-periodic system into the frequency domain results in inter-frequency coupling \citep{islam2024identification}. We use a new formulation of harmonic resolvent analysis for bilinear systems that ensures that all inter-frequency coupling takes the form of triadic interactions, simplifying their interpretation. The analysis, therefore, captures the energy redistribution across harmonics driven by triadic coupling between the screech mode and higher-order frequencies, such as those at twice and thrice the screech frequency. The components of the optimal harmonic resolvent mode at harmonics of the screech frequency are consistent with experimental observations \citep{powell1953mechanism, tam2014harmonics} that have not been previously predicted by global models.

Fourth, we leverage our bilinear formulation of harmonic resolvent analysis to study the impact of the screech mode's nonlinear self-interaction on other fluctuations in the jet.  Rather than assume the forcing applied to the harmonic resolvent operator to be unknown and subsequently optimize it as in standard resolvent and harmonic resolvent analyses, our analysis directly computes the response of the jet to the actual nonlinear self-interaction term.  We show that the forcing provided by this nonlinear self-interaction of the screech mode, along with its triadic interactions with other frequencies embedded in the harmonic resolvent operator, is sufficient to explain the redistribution of energy to other frequencies without appealing to an unknown forcing from background turbulence.

Computing resolvent and harmonic resolvent modes can be challenging, as it involves applying the linear operator to test vectors iteratively until the forcing and response modes converge. This can be done in the frequency domain using either the Arnoldi method or randomized singular-value decomposition (RSVD), with the former generally outperforming the latter \citep{Ribeiroetal20}. A key computational bottleneck in both approaches is the need for a lower-upper (LU) decomposition of the operator in the frequency domain. We refer to this approach, pairing RSVD and LU decomposition, as the RSVD-LU algorithm. Alternatively, a second approach obtains the action of the resolvent or harmonic resolvent operator on a forcing vector by integrating the linearized equations in the time domain before applying a Fourier transform, eliminating the need for an LU decomposition \citep{Monokrousosetal10, Martinietal21, Farghadanetal21}. The RSVD-\Dt algorithm combines this time-stepping approach with RSVD \citep{farghadan2024scalable}. A key advantage of RSVD-\Dt is its scalability with system dimension, both in terms of memory consumption and CPU time, making it ideal for the large frequency-coupled systems arising in harmonic resolvent analysis \citep{Farghadanetal24}.


The structure of this paper is as follows. In \S\ref{sec:experiment}, we outline the experimental setup and subsequent data processing used to obtain the mean flow and isolate the screech mode. In \S\ref{sec:resolvent} and \S\ref{sec:harmonic}, we provide an overview of the global eigendecomposition, resolvent, and harmonic analyses for a bilinear system, along with their computational methods. In \S\ref{sec:nonlinear}, we introduce the nonlinear forcing analysis enabled by our bilinear formulation of harmonic resolvent analysis.  We report our results for each of these analyses in \S\ref{sec:results}, and we summarize our findings in \S\ref{sec:conclusion}.

\section{Database and methodology}

\subsection{Experimental database} 
\label{sec:experiment}

We consider a screeching jet resonating in an axisymmetric ($m = 0$) mode, characterized by an ideally expanded jet Mach number of $M_j = \frac{U_j}{a_j} = 1.12$ and an acoustic Mach number of $M_a = \frac{U_j}{a_{\infty}} = 1$, where $U_j$ is the ideally expanded jet velocity at the nozzle exit, $a_j$ is the speed of sound at the nozzle exit, and $a_{\infty}$ is the ambient speed of sound. The jet is not heated, so a nearly isentropic temperature profile is expected. The Reynolds number at this Mach number is $\Rey=\frac{\rho_j U_j D_j}{\mu_j}\approx3.41\times10^5$, where $\rho_j$ is the ideally expanded jet density, $D_j$ is the ideally expanded nozzle diameter, and $\mu_j$ is the ideally expanded jet viscosity found using the Sutherland gaseous viscosity model. The screech frequency for this jet, obtained from acoustic measurements, corresponds to Strouhal number $St = \frac{\omega D_j}{2\pi U_j} = 0.714$, where $\omega$ is the dimensional frequency. 

The experimental data analyzed in this work are obtained from non-time-resolved particle-image velocimetry (PIV) measurements performed in the Supersonic-Jet Anechoic Facility (SJAF) at Monash University. This modular facility has been described in several previous works \citep{beekman2024acoustic,nogueira2024screech} and has been used to evaluate screech characteristics in jets of various geometries. A round converging nozzle of diameter $D=10$ mm was printed from Phrozen Aqua 8K resin on a Phrozen Sonic Mighty 8K resin printer and attached to a plenum. PIV and acoustic experiments were performed for the jet conditions described above. A 12-bit Imperx B6640M camera with a resolution of 6600 px $\times$ 4400 px was used in conjunction with a $300$ mm Nikkor lens to obtain $3,000$ single-exposure image pairs with a spatial resolution of $16.4$ $\umu$m/px at an acquisition rate of $1.66$ Hz. Extension rings were used to provide the appropriate field of view, resulting in a total extension of $45$ mm. The light sheet was generated by a Quantel Evergreen dual-cavity pulsed Nd:YAG laser, with a wavelength of $532$ nm and a thickness of $1$ mm. Atomized paraffin particles were used to seed the flow in the settling chamber, with particle sizes of approximately 0.6-0.8 $\umu$m. Within the measurement region of interest, light-based reflections were not observed.

The particle snapshot pairs were pre- and post-processed in MATLAB using PIVLab, an open-source tool that leverages a multi-grid algorithm \citep{soria1996investigation}, with an initial window size of $128\times128$ px and a final window size of $32\times32$ px. The raw images are pre-processed using contrast-limited adaptive histogram equalization (CLAHE) and a Wiener adaptive noise filter with a kernel size of 3 px. A local normalized median threshold of 2.5 is used for vector validation, and individual velocity vectors greater than 5 standard deviations from the temporal mean were replaced with the mean value.

\begin{figure}
\centering
\includegraphics[width=\textwidth]{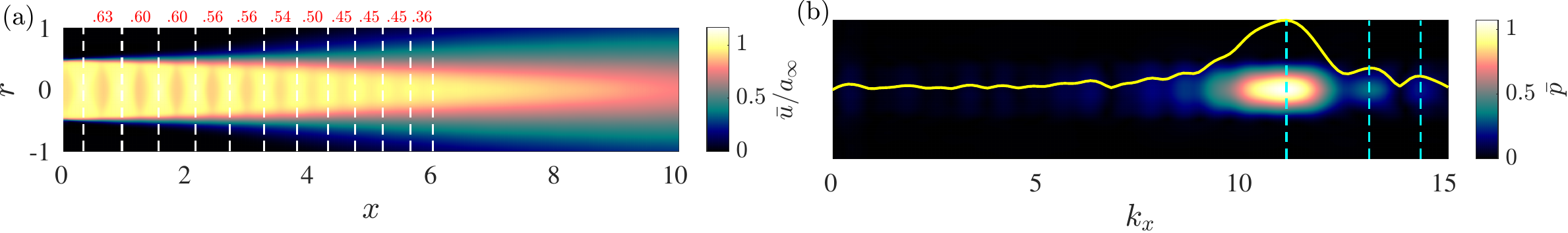}
\caption{Analysis of the shock-cell structure in the mean flow: (a) Mean streamwise velocity showing weak shock cells, with dashed lines indicating the shock cell spacing. (b) Wavenumber spectrum of the mean pressure, with the yellow line representing the normalized wavenumber along the centerline at $r = 0$, and cyan dashed lines mark the local maxima of the spectrum along the centerline.}
\label{fig:screechWaveNumShock}
\end{figure}

These PIV data are used to compute the mean flow for subsequent linear analyses.  The mean streamwise and radial velocity fields are directly provided by the time-averaged PIV data, revealing the presence of weak shock cells within the flow (see figure \ref{fig:screechWaveNumShock}(a) for a visualization).  Other mean quantities required for the linearization, such as pressure and specific volume, are derived by employing the Crocco-Busemann relation \citep{Busemann31, Crocco32} in conjunction with the equation of state. 

To isolate the coherent structures associated with the screech phenomenon, snapshot proper orthogonal decomposition (POD) is applied to the PIV velocity fields \citep{Lumley67, Sirovich87_1, Berkoozetal93}. This method decomposes the flow into orthogonal spatial modes ranked by energy content, effectively capturing dominant flow features. The leading POD mode pair, with streamwise velocity fluctuations that are symmetric about the jet centerline for axisymmetric screech modes, encapsulates the periodic fluctuations at the screech frequency $St = 0.714$, corresponding to the aeroacoustic feedback loop \citep{Edgingtonetal18}. These modes, which will be shown later, represent the spatial structures of upstream- and downstream-traveling waves, such as KH wavepackets and upstream-propagating acoustic guided jet modes, providing a point of comparison for the subsequent analyses. Because the PIV snapshots are not time-resolved, we identify the leading POD pair as the screech mode at $St_s = 0.714$ by matching it to the acoustic spectrum, following the convention used in prior experimentally driven modal analyses of screeching jets \citep{Edgingtonetal18, Edgingtonetal21}; the leading POD pair is the appropriate object for this identification because it captures the dominant cyclic structure in the snapshot ensemble \citep{Oberleithneretal11}.

\subsection{Linear mean flow analysis of a bilinear system} \label{sec:resolvent}

The compressible Navier-Stokes equations, which govern screeching jets, can be written in numerous equivalent forms depending on the choice of the thermodynamic variables included in the definition of the state vector.  In our analyses presented in \S\ref{sec:results}, we use the bilinear form obtained by choosing specific volume and pressure as our two thermodynamic variables.  This choice will prove advantageous for the careful study of mode interactions; the rationale and the explicit state-vector form are given in \S\ref{sec:setup}.  In this case, the Navier-Stokes equations can be expressed as \citep{Towne16}
\begin{equation}
\frac{\partial \bm q}{\partial t} = \boldsymbol{\mathcal{B}}(\bm q, \bm q),
\label{eqn:bilinear}
\end{equation}
where $\bm q$ denotes the state vector of flow variables, and $\boldsymbol{\mathcal{B}}$ is the bilinear Navier-Stokes operator capturing the system's quadratic nonlinearity. 

To investigate fluctuations about the mean, we begin with the Reynolds decomposition
\begin{equation}
\bm q(\bm x, t) = \bar{\bm q}(\bm x) + \bm q'(\bm x, t),
\end{equation}
where $\bm q'$ is the fluctuation to the time-averaged mean $\bar{\bm q}$.

Substituting this decomposition into \eqref{eqn:bilinear} and leveraging the bilinear nature of $\boldsymbol{\mathcal{B}}$ yields
\begin{equation}
\frac{\partial \bm q'}{\partial t} = \boldsymbol{\mathcal{B}}(\bar{\bm q}, \bm q') + \boldsymbol{\mathcal{B}}(\bm q', \bar{\bm q}) + \boldsymbol{\mathcal{B}}(\bar{\bm q}, \bar{\bm q}) + \boldsymbol{\mathcal{B}}(\bm q', \bm q').
\label{eq:LNS_bilinearForm}
\end{equation}
By defining 
\begin{equation}
\bm A(\bar{\bm q}) \bm q' = \boldsymbol{\mathcal{B}}(\bar{\bm q}, \bm q') + \boldsymbol{\mathcal{B}}(\bm q', \bar{\bm q})
\end{equation}
to contain the terms that are linear in the fluctuation $\bm q'$ and 
\begin{equation}
\bm B \bm f = \boldsymbol{\mathcal{B}}(\bar{\bm q}, \bar{\bm q}) + \boldsymbol{\mathcal{B}}(\bm q', \bm q')
\end{equation}
to contain the constant and nonlinear terms, (\ref{eq:LNS_bilinearForm}) can be equivalently expressed in the more familiar form
\begin{equation}
\frac{\partial \bm q'}{\partial t} = \bm A(\bar{\bm q}) \bm q' + \bm B \bm f.
\label{eqn:linear}
\end{equation}
Here, $\bm A(\bar{\bm q})$ is the linearized Navier-Stokes operator governing the linear dynamics, $\bm B$ maps the forcing term $\bm f$ to the state, and $\bm f$ is interpreted as the effective forcing on the linear dynamics, treated as unknown (i.e., assumed independent of the fluctuation $\bm q'$) in the resolvent optimization following the standard convention. The constant component $\boldsymbol{\mathcal{B}}(\bar{\bm q}, \bar{\bm q})$ within $\bm B \bm f$ is the residual forcing that sustains the chosen base $\bar{\bm q}$ in the steady linear system. An output of interest can be extracted from the state as
\begin{equation}
\bm y' = \bm C \bm q'.
\label{eqn:output}
\end{equation}
This framework provides a foundation for both eigenvalue decomposition and resolvent analysis, as described below.

The long-time asymptotic response of (\ref{eqn:linear}) to an initial disturbance in the absence of forcing $\bm f$ is determined by the eigenvalues of $\bm A$. While the terms represented by $\bm f$ are not zero in real flows, eigendecomposition is an effective tool for identifying resonant phenomena such as screech, which manifests as lightly damped discrete eigenvalues.  Assuming a normal mode ansatz for the fluctuation, 
\begin{equation}
    \bm q'(\bm x, t) = \hat{\bm q}(\bm x)e^{-\text{i} \omega t}, 
\end{equation}
and neglecting external forcing in \eqref{eqn:linear}, we obtain the eigenvalue problem
\begin{equation} 
\bm A \hat{\bm{q}} = -\text{i} \omega \hat{\bm{q}},
\end{equation}
where $\text{i} = \sqrt{-1}$ is the imaginary unit and $\omega = \omega_r + \text{i} \omega_i$ represents the complex frequency.  The real part $\omega_r$ corresponds to the oscillation frequency, while the imaginary part $\omega_i$ determines stability: $\omega_i > 0$ signifies exponential growth and instability, $\omega_i < 0$ indicates temporal decay and stability, and $\omega_i = 0$ denotes neutral stability.

To analyze the steady-state response of the system, \eqref{eqn:linear} and \eqref{eqn:output} are transformed into the frequency domain via a Fourier transform, giving
\begin{equation}
\label{eq:resolvent_transfer_fcn}
\hat{\bm y}(\omega) =  \bm C \bm R(\omega) \bm B \hat{\bm f}(\omega) , 
\end{equation}
where 
\begin{equation}
\bm R(\omega) = (-\text{i}\omega \bm I - \bm A)^{-1} 
\end{equation}
is the resolvent operator and (\ref{eq:resolvent_transfer_fcn})
maps the Fourier-transformed forcing $\hat{\bm f}$ to the transformed output $\hat{\bm y}$.  Here, $\bm I$ is the identity matrix.  Since the forcing $\hat{\bm f}$ is usually unknown, the goal of a typical resolvent analysis is to identify the optimal forcing $\hat{\bm f}$ that maximizes the system's response $\hat{\bm y}$.  If the gain
\begin{equation}
\sigma^2 = \frac{||\hat{\bm y}||_y^2}{||\hat{\bm f}||_f^2}
\end{equation}
is large, then the corresponding forcing and response are likely to be relevant regardless of the details of the true forcing.  The norm
\begin{equation}
    ||\bm y||_y^2 = \bm y^* \bm W_y \bm y
\end{equation}
measures the energy of a given response $\bm y$, where $\bm W_y$ is a weight matrix used to define the desired norm and $(\cdot)^*$ denotes the complex conjugate transpose. Note that the input norm $||\cdot||_f$ may differ from the output norm $||\cdot||_y$. 

The optimal gain and the corresponding modes are obtained via singular value decomposition (SVD) of the resolvent operator with appropriate weighting to ensure optimality in the desired norms.  When the weights are diagonal matrices, this is usually accomplished by defining the weighted resolvent operator \citep{Towneetal18}
\begin{equation}
\tilde{\bm R} = \bm W_q^{1/2} \bm C (-\text{i}\omega \bm I - \bm A)^{-1} \bm B \bm W_f^{-1/2}. 
\label{eqn:mod_Res}
\end{equation}
 When the weights are not diagonal, as is the case in this study (see \S\ref{sec:setup} and Appendix~\ref{appA}), it is more computationally efficient to compute their Cholesky decomposition rather than the inverses in~(\ref{eqn:mod_Res}). Following the methodology outlined by \citet{Herrmannetal21}, and utilizing the Cholesky factorizations $\bm W_y = \bm F_y^* \bm F_y$ and $\bm W_f = \bm F_f^* \bm F_f$, which are guaranteed by the positive definiteness of the weight matrices, the appropriate weighted resolvent operator is
\begin{equation}
\tilde{\bm R} = \bm F_y \bm C (-\text{i}\omega \bm I - \bm A)^{-1} \bm B \bm F_f^{-1}.
\end{equation}
Computing the SVD
\begin{equation}
\tilde{\bm R} = \tilde{\bm U}^R \boldsymbol{\Sigma}^R (\tilde{\bm V}^R)^*,
\end{equation}
an ordered set of optimal gains is given by the singular values contained in the diagonal matrix $\boldsymbol{\Sigma}^R = \text{diag}(\sigma_1^R, \sigma_2^R, \ldots)$, and the corresponding forcing and response modes are contained in the columns of the matrices
\begin{equation}
\begin{gathered}
\bm U^R = \bm F_y^{-1} \tilde{\bm U}^R = [\hat{\bm q}_1^R, \hat{\bm q}_2^R, \ldots], \\
\bm V^R = \bm F_f^{-1} \tilde{\bm V}^R = [\hat{\bm f}_1^R, \hat{\bm f}_2^R, \ldots].
\end{gathered}
\label{eqn:UV_rec2}
\end{equation}

\subsection{Harmonic resolvent analysis of a bilinear system} \label{sec:harmonic}

Harmonic resolvent analysis extends the traditional resolvent framework to systems with periodic base flows, providing deeper insight into cross-frequency interactions \citep{Padovanetal20, PadovanRowley22}. Here, we describe the framework for the bilinear system introduced in the previous section, which we show has critical advantages over considering general nonlinearities.

As in \S\ref{sec:resolvent}, the first step is to apply Reynolds decomposition. However, in this case, the state vector is decomposed into a time-averaged mean $\bar{\bm q}$, a periodic base component $\bm q_T(t) = \bm q_T(t+T)$, and fluctuations $\bm q'(t)$, i.e.,
\begin{equation}
\bm q(\bm x, t) = \bar{\bm q}(\bm x) + \bm q_T(\bm x, t) + \bm q'(\bm x, t).
\label{eqn:Reynolds}
\end{equation}
Substituting this decomposition into \eqref{eqn:bilinear} yields
\begin{multline}
\frac{\partial \bm q'}{\partial t} = \boldsymbol{\mathcal{B}}(\bar{\bm q} + \bm q_T, \bm q') + \boldsymbol{\mathcal{B}}(\bm q', \bar{\bm q} + \bm q_T) + \boldsymbol{\mathcal{B}}(\bar{\bm q}, \bar{\bm q}) + \boldsymbol{\mathcal{B}}(\bm q_T, \bm q_T) \\ + \left(-\frac{\partial \bm q_T}{\partial t} + \boldsymbol{\mathcal{B}}(\bm q_T, \bar{\bm q}) + \boldsymbol{\mathcal{B}}(\bar{\bm q}, \bm q_T)\right)   + \boldsymbol{\mathcal{B}}(\bm q', \bm q').
\label{eqn:bilinear2}
\end{multline}

Noting that only the first two terms on the right-hand side are linear with respect to the fluctuation $\bm q'$, (\ref{eqn:bilinear2}) can be written in compact form as

\begin{equation}
\frac{\partial \bm q'}{\partial t} = \bm A_T(t) \bm q' + \bm B \bm f,
\label{eqn:linsys_harmonic}
\end{equation}
where 
\begin{equation}
\bm A_T(t) \bm q'= \boldsymbol{\mathcal{B}}(\bar{\bm q} + \bm q_T, \bm q') + \boldsymbol{\mathcal{B}}(\bm q', \bar{\bm q} + \bm q_T)
\label{eqn:linsys_harmonic_linearTerms}
\end{equation}
and 
\begin{equation}
\bm B \bm f = \boldsymbol{\mathcal{B}}(\bar{\bm q}, \bar{\bm q}) + \boldsymbol{\mathcal{B}}(\bm q_T, \bm q_T) - \bm f_T + \boldsymbol{\mathcal{B}}(\bm q', \bm q').
\label{eqn:nonlinear_terms_harmonic}
\end{equation}
Here, $\bm A_T(t) = \bm A_T(t + T)$ is a time-periodic linear operator.  The base frequency is denoted by $\omega_f$, with the fundamental period defined as $T = 2\pi/\omega_f$. By construction, interactions between fluctuation $\bm q'$ and both $\bar{\bm q}$ and $\bm q_T$ are retained in the linear operator.  Notably, the bilinear form guarantees that $\bm A_T(t)$ contains the same frequencies as the periodic base component $\bm q_T(t)$, unlike harmonic resolvent analyses of general nonlinear systems.  The term $ \bm f_T = \frac{\partial \bm q_T}{\partial t} - \bm A(\bar{\bm q}) \bm q_T$ can be interpreted as the forcing that would need to be applied to the steady linear system from \S\ref{sec:resolvent} to generate the periodic base component $\bm q_T(t)$. We emphasize that this is a mathematical interpretation: $\bm f_T$ is the residual that would close the steady linear system from \S\ref{sec:resolvent} if $\bm q_T$ were treated as a fluctuation generated within that system, not a physical forcing that drives $\bm q_T$ in the full nonlinear dynamics. As in \S\ref{sec:resolvent}, $\bm B \bm f$ collects the nonlinear contributions: $\boldsymbol{\mathcal{B}}(\bar{\bm q}, \bar{\bm q})$, $\boldsymbol{\mathcal{B}}(\bm q_T, \bm q_T)$ (now time-periodic), the residual $-\bm f_T$, and the internal self-interaction $\boldsymbol{\mathcal{B}}(\bm q', \bm q')$.  Harmonic resolvent analysis treats $\bm f$ as an unknown nonlinear forcing and seeks the optimal forcing with the largest gain, which, analogous to the leading resolvent mode, is likely to appear in the solution for an arbitrary forcing.  

The periodic nature of the base flow ensures that both $\bm A_T(t)$ and $\bm q'(t)$ can be expanded in Fourier series as
\begin{equation}
\begin{gathered}
\bm A_T(t) = \sum_{j = -\infty}^{\infty} \hat{\bm A}_{p, j} e^{-\text{i} j\omega_f t}, 
\\
\bm q'(t) = \sum_{j = -\infty}^{\infty} \hat{\bm q}_j e^{-\text{i} j\omega_f t}, 
\label{eqn:periodic_A_q}
\end{gathered}
\end{equation}
where $(\cdot)_j$ denotes the $j^{th}$ harmonic of $\omega_f$. This transformation reformulates the system in the frequency domain, naturally embedding interactions across harmonics. The resulting set of coupled equations for the Fourier ansatz can be written compactly as \citep{Padovanetal20, Farghadanetal24}
\begin{equation}
\begin{gathered}
\bm T \hat{\bm q} = \bm B \hat{\bm f},
\\
\hat{\bm y} = \bm C \hat{\bm q}.
\end{gathered}
\label{eqn:Hres_eqn1}
\end{equation}

Here, $\hat{\bm q}$, $\hat{\bm f}$, and $\hat{\bm y}$ are infinite-dimensional vectors containing the Fourier modes of the state, forcing, and output, respectively, for all harmonics of $\omega_f$.  Similarly, $\bm T$ is an infinite-dimensional block-structured operator that encodes harmonic interactions, with diagonal terms involving the resolvent operator at each harmonic frequency and off-diagonal terms capturing coupling induced by the time-periodic base flow.  In practice, these variables are made finite-dimensional by truncating the Fourier series to some maximum frequency.  The bilinear form of $\boldsymbol{\mathcal{B}}$ ensures a minimal set of triadic interactions, which both minimizes the impact of truncation and aids in the interpretation of results.

The harmonic resolvent operator $\bm H = \bm T^{-1}$ provides a direct mapping from harmonic forcing to the resulting responses,
\begin{equation}
\hat{\bm y} = \bm C \bm H \bm B \hat{\bm f}.
\label{eqn:Hres_eqn2}
\end{equation}
This operator inherently incorporates all harmonics of the base frequency, distinguishing harmonic resolvent analysis from traditional approaches that treat each frequency independently. Similar to resolvent analysis, the weighted harmonic resolvent operator is employed to account for the appropriate norms. The SVD is computed as
\begin{equation}
\tilde{\bm H} = \bm F_y \bm C \bm H \bm B \bm F_f^{-1} = \tilde{\bm U}^H \boldsymbol{\Sigma}^H (\tilde{\bm V}^H)^*,
\end{equation}
where $\boldsymbol{\Sigma}^H = \text{diag}(\sigma_1^H, \sigma_2^H, \ldots)$ contains the optimal harmonic resolvent gains, and the modes are recovered as
\begin{equation}
\begin{gathered}
\bm U^H = \bm F_q^{-1} \tilde{\bm U}^H = [\hat{\bm q}_1^H, \hat{\bm q}_2^H, \ldots], \\
\bm V^H = \bm F_f^{-1} \tilde{\bm V}^H = [\hat{\bm f}_1^H, \hat{\bm f}_2^H, \ldots].
\end{gathered}
\end{equation}

From a computational standpoint, harmonic resolvent analysis poses unique challenges due to the increased dimensionality introduced by harmonic coupling.  The RSVD-LU algorithm from \citet{Padovanetal20} is not feasible here due to its high memory requirements, which exceed the available capacity on Michigan's Great Lakes cluster for the screeching jet configuration. To make the computations tractable, we employ the RSVD-\Dt algorithm as described in \citet{Farghadanetal24}, which uses RSVD along with time-stepping to approximate the SVD of $\tilde{\bm H}$.

\subsection{Nonlinear forcing in harmonic resolvent analysis} \label{sec:nonlinear}

Harmonic resolvent analysis, as introduced by \citet{Padovanetal20} and described in \S\ref{sec:harmonic}, treats $\bm f$ as an unknown nonlinear forcing and seeks optimal forcing-response pairs.  Our bilinear formulation makes it possible to isolate the specific, known component of this nonlinear forcing that is contributed by the periodic base itself, associated with a given choice of the periodic base component $\bm q_T(t)$.  The term of particular interest is 
\begin{equation}
\hat{\bm f}_{nl} = \boldsymbol{\mathcal{B}}(\bm q_T, \bm q_T),
\label{eqn:nonlinearF}
\end{equation}
which explicitly represents nonlinear self-interactions generated by the periodic part of the base flow.  The response of the fluctuation $\bm q'$ to this nonlinearity is
\begin{equation}
\hat{\bm q}_{nl} = {\bm H} \hat{\bm f}_{nl}.
\label{eqn:nonlinearY}
\end{equation}

In this paper, $\bm q_T(t)$ represents the screech mode.  Accordingly, (\ref{eqn:nonlinearY}) enables us to study how the nonlinear self-interaction of the screech mode generates other perturbations in the jet.  More broadly, our bilinear formulation complements the standard harmonic-resolvent framework: rather than treating the entire forcing as unknown and optimizing over it, we use the harmonic-resolvent operator $\bm H$ directly to map the specific known portion of the nonlinear forcing contributed by the periodic base to its response. 

\section{Results} \label{sec:results}

To fully characterize the screech phenomenon, global stability and resolvent analyses are employed to compute the screech mode, and the results are compared with the POD modes derived from the PIV data. Building on this foundation, the triadic and nonlinear interactions are explored using harmonic resolvent analysis and our nonlinear extension.

\subsection{Building the linear operator} \label{sec:setup}

In this study, the global state vector, compactly represented as 
\begin{equation}
\bm q = [\xi, u, v, w, p]^T(x, r, \theta, t), 
\label{eqn:statevariables}
\end{equation}
includes the specific volume $\xi$, the velocity components $u$, $v$, and $w$ in the streamwise, radial, and azimuthal directions, respectively, and the pressure $p$. Specifically, $1/\bar \rho$, $a_{\infty}$, and $\bar \rho a_{\infty}^2$ are used for the nondimensionalization of specific volume, velocities, and pressure, respectively, and the perfectly expanded nozzle diameter $D_j$ is used to nondimensionalize spatial coordinates. The rationale for the specific-volume/pressure choice introduced in \S\ref{sec:resolvent} is that the ideal-gas law takes the bilinear form $T = \gamma p \xi$, which together with the specific-volume substitution $\rho = 1/\xi$ renders every nonlinear term on the right-hand side of the Navier-Stokes equations a product of exactly two components of $\bm q$; see \citet{Towne16} for additional details. 

The operator $\bm A$ is constructed by linearizing the compressible Navier-Stokes equations around the PIV-derived base flow. Using a normal mode ansatz in the azimuthal and temporal dimensions, the state vector can be expressed as
\begin{equation}
\bm{q}(x, r, \theta, t) = \hat{\bm{q}}(x, r)e^{-\text{i}\omega t + \text{i}m\theta},
\label{eqn:screechState}
\end{equation}
where $\hat{\bm{q}}(x, r)$ is the spatially varying amplitude of the mode, $m$ is the azimuthal wavenumber, and $\omega$ is the angular frequency. Thus, each mode is characterized by a unique pair $(m, \omega)$, representing its azimuthal and temporal behavior. In this study, we focus exclusively on the axisymmetric mode ($m = 0$), where the screech phenomenon occurs for the current jet operating conditions.

The computational domain of interest spans $x \times r \in [0, 15] \times [0, 3]$. The experimental mean flow is extended downstream of the $x = 10$ boundary of the PIV window to $15$ by imposing a self-similar velocity profile from \citet[Sec.~5.2]{Pope00}, following prior resolvent and stability analyses of round turbulent jets \citep{lesshafft2019resolvent, Edgingtonetal21}. Sponge layers are applied at the inlet, outlet, and far-field boundaries to minimize numerical reflections. The boundary conditions within the sponge region are designed to effectively absorb outgoing waves, thereby reducing reflections and numerical artifacts \citep{Mani12, Schmidtetal17}. These sponge layers have been consistently employed in previous numerical studies \citep{Edgingtonetal21, Prasadetal22, GomezMcKeon25}. The computational domain, including the sponge layers, is discretized with $N_x \times N_r = 700 \times 300$ grid points in the streamwise and radial directions, respectively. A finer grid resolution is adopted in the core region, where smaller flow structures are anticipated, specifically within $x \times r \in [0, 4] \times [0, 0.7]$, where a non-uniform grid with $350 \times 65$ points is used in the streamwise and radial directions, respectively. Spatial derivatives are computed using fourth-order summation-by-parts finite-difference schemes \citep{MattssonNordstrom04}. We verified mesh independence by also computing the leading resolvent modes and gains on a coarser $(N_x \times N_r) = (400, 200)$ grid; differences are small and do not modify the conclusions of the analysis, consistent with the large-scale nature of the dominant Kelvin-Helmholtz wavepackets and guided jet modes that are the focus of the present analysis. This grid resolution is also similar to those used in prior global resolvent studies of round jets \citep{Schmidtetal18, Pickeringetal21, farghadan2024scalable}. The linearized Navier-Stokes operator employs a reduced Reynolds number of 1000 to approximate the influence of unmodeled Reynolds stresses \citep{ReynoldsHussain72, Schmidtetal18, Pickeringetal21}, consistent with the approach in our earlier work \citep{farghadan2024scalable}. In particular, \citet{Pickeringetal21} found $Re = 1000$ to be close to the optimal eddy-viscosity for a turbulent round jet over a range of frequencies.  Moreover, the close agreement between the leading resolvent mode and the leading POD mode at the screech frequency (figure~\ref{fig:screechPODComp}) provides an a posteriori check that this choice is reasonable for the present jet.

The norms required for resolvent and harmonic resolvent analyses are defined using Chu's compressible energy norm \citep{Chuetal65} for both forcing and response modes. The weight matrix $\bm W = \bm W_f = \bm W_q$, derived in Appendix \ref{appA} for our specific-volume-pressure formulation, compensates for the non-uniform grid when computing Chu's energy norm for the non-dimensionalized flow state $\bm q$. As a result, we define the norm as $|| \cdot || = || \cdot ||_y = || \cdot ||_f$, ensuring consistency across the input and output modes. The output and input matrices, $\bm C$ and $\bm B$, are defined over the domains $x \times r \in [0, 15] \times [0, 3]$ and $x \times r \in [0, 15] \times [0, 1]$, respectively, excluding the sponge region. These matrices are applied across all analyses, including both resolvent and harmonic resolvent analyses.

\subsection{Eigenspectra analysis} \label{sec:eig}

\begin{figure}
\centering
\includegraphics[scale=0.4]{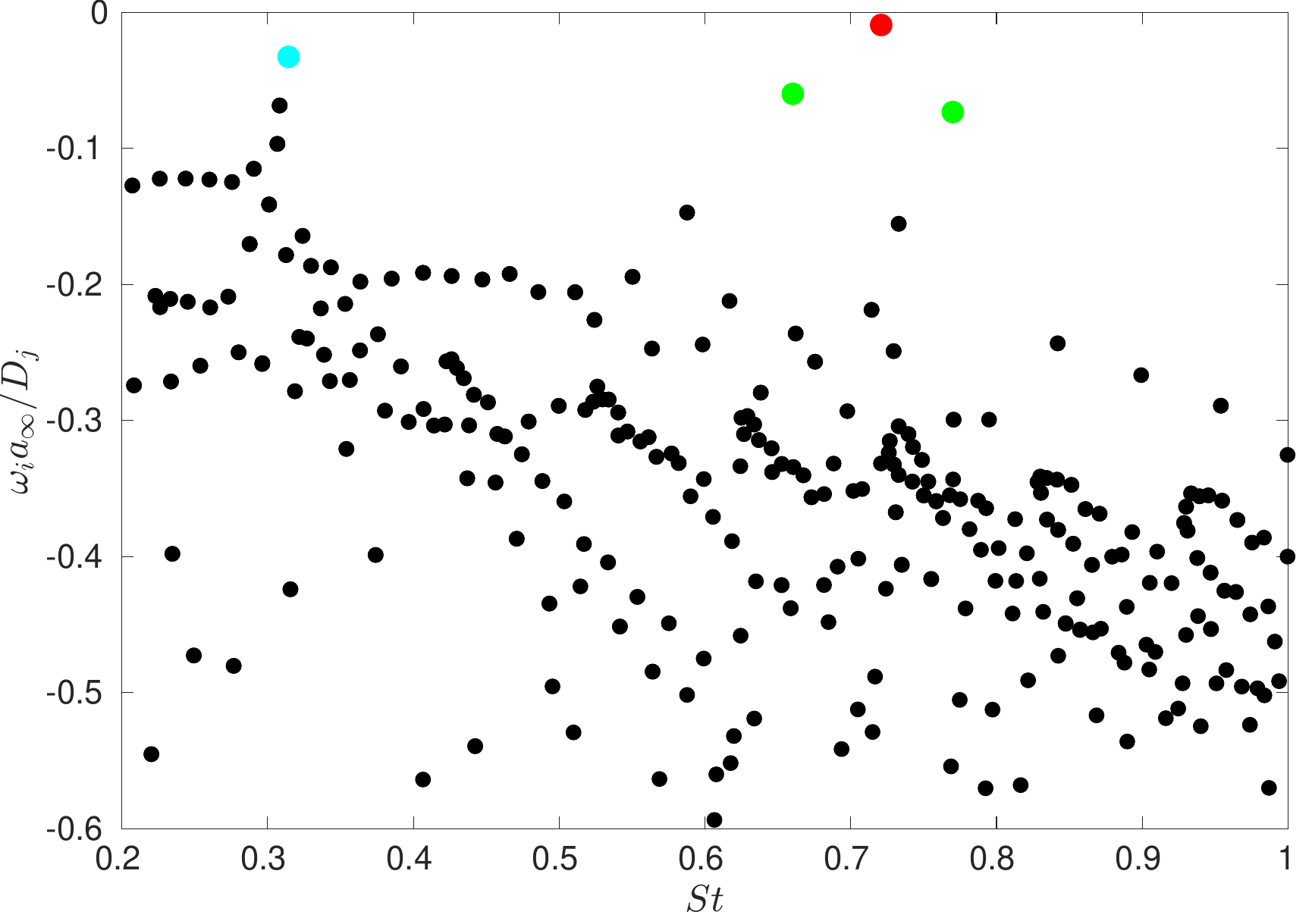}
\caption{Eigenspectrum of the linearized operator around the mean flow. The red eigenvalue is the least-stable mode, which closely matches the observed screech frequency. We show that the two green eigenvalues are additional candidate screech modes created by interactions with different shock-cell wavenumbers. The blue eigenvalue is caused by resonance between upstream- and downstream-traveling acoustic duct modes.}
\label{fig:screechEigSpectrum}
\end{figure}

We begin our analysis of the screech phenomenon by computing the global eigenspectra of $\bm A$.  Previous studies have shown that the screech mode manifests as a lightly-damped, isolated eigenmode in similar jets \citep{beneddine2015global, Edgingtonetal21}. 

The eigenvalue spectrum of $\bm A$ for our jet is presented in figure \ref{fig:screechEigSpectrum} for $St \in [0.2, 1]$. The spectrum consists of several branches and a number of discrete modes. The screech mode, in particular, is expected to manifest at close to the experimentally observed screech frequency. The reported screech frequency is $St = 0.714$, and our analysis identifies the least stable mode (highlighted in red) at $St = 0.72$, corresponding to a relative discrepancy of less than 1\%. In addition to the primary screech mode, three other lightly damped discrete eigenvalues (colored cyan and green in the figure) are observed; these modes will be analyzed shortly.

\begin{figure}
\centering
\includegraphics[width=\textwidth]{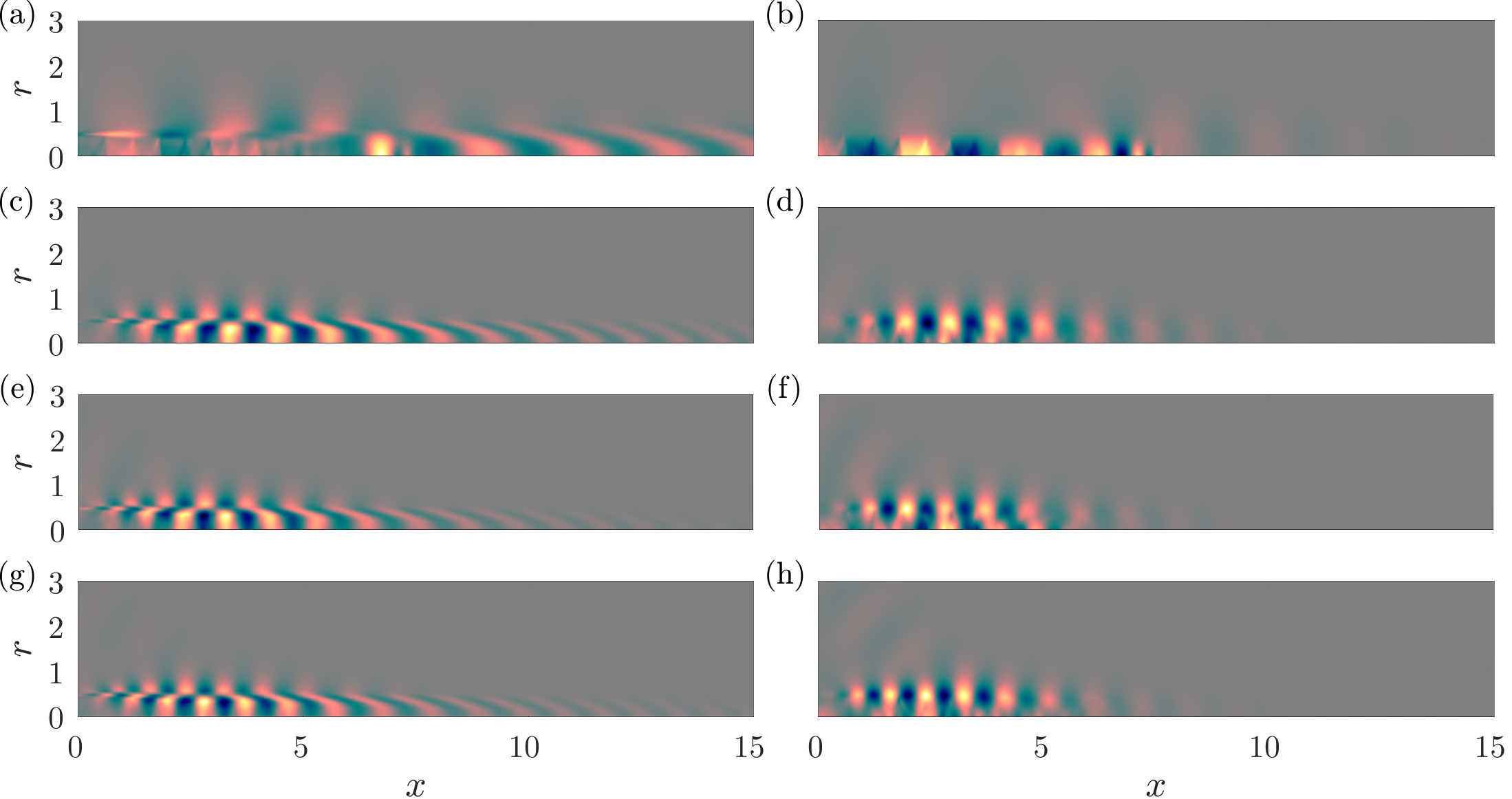}
\caption{The eigenmodes associated with isolated eigenvalues of the screeching jet. Panels (a, c, e, g) show the streamwise velocity, and panels (b, d, f, h) show the pressure at frequencies of 0.31 (cyan), 0.66 (green), 0.72 (red), and 0.77 (green), respectively. The color coding in parentheses corresponds to the spectrum shown in figure \ref{fig:screechEigSpectrum}.}
\label{fig:screechEigModes}
\end{figure}

Figure \ref{fig:screechEigModes} illustrates the real part of the streamwise velocity and pressure components of the eigenmodes corresponding to the four aforementioned discrete eigenvalues. For all subsequent figures of such modes, we show only the real parts, though the modes are inherently complex. For the mode at $St = 0.72$, the pressure field reveals a KH wavepacket localized near the shear layer ($r \approx 1/2$) and a guided jet mode concentrated within the jet core ($r < 1/2$), consistent with the interaction between a downstream-traveling KH wavepacket and an upstream-traveling guided jet mode within the screech resonance cycle, as detailed by \citet{Edgingtonetal21}. These characteristics, with KH instabilities dominating in the shear layer and guided modes concentrated within the core, confirm that this mode is a screech mode. The modes at $St = 0.66$ and $St = 0.77$ exhibit similar features, with wavepackets differing in length and wavenumber due to the change in frequency. We will show in \S\ref{sec:res} that these discrete modes arise from interactions between KH instabilities and suboptimal wavenumbers of the shock-cell structure \citep{nogueira2022closure}.  Finally, the mode at $St = 0.31$ contains both high and low-wavenumber structures in the jet core but lacks a clear KH wavepacket.  This structure, as well as the frequency of the eigenvalue, is consistent with resonance between duct-like and guided jet modes, as described by \citet{towne2017acoustic}. The localized peak at the centerline near $x \approx 7$ in figure~\ref{fig:screechEigModes}(a) reflects the same physics: the contraction of the potential core produces a turning point near that streamwise location, where upstream- and downstream-traveling duct modes resonate \citep{Schmidtetal17}. From this figure onwards, forcing and response modes are normalized to unit Chu energy and colorbars are accordingly omitted, since the absolute color values are normalization-dependent; the sign and spatial structure of each mode (and, for wavenumber spectra of normalized modes, the relative magnitude and peak locations) carry the physical interpretation.

\begin{figure}
\centering
\includegraphics[width=\textwidth]{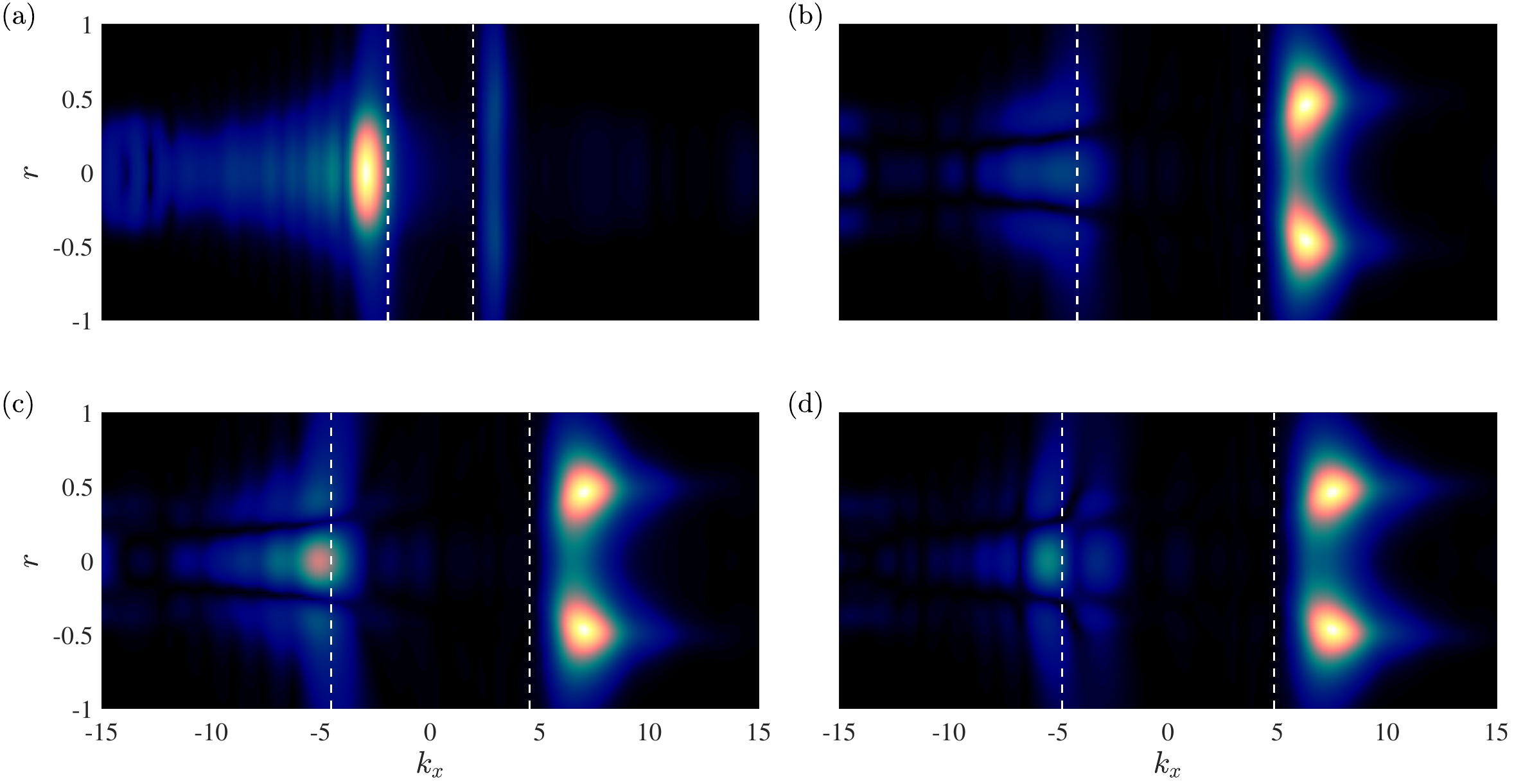}
\caption{Wavenumber spectra of the pressure component of the eigenmodes at (a-d) $St = 0.31$, $0.66$, $0.72$, and $0.77$, respectively. White dashed lines indicate the acoustic wavenumbers $k_x = \pm \omega$.}
\label{fig:screechWaveNumEigModes}
\end{figure}

To further investigate the dynamics of these four discrete global modes, we perform a wavenumber transform in the streamwise direction. In what follows, we use $k_{kh}$, $k_{gj}$, and $k_s$ to denote the streamwise wavenumbers of the Kelvin-Helmholtz wavepacket, the guided jet mode, and the shock cells, respectively; their relationship is given formally in~\eqref{eqn:WaveNum2}. Figure \ref{fig:screechWaveNumEigModes} presents the wavenumber spectra of the pressure component of the eigenmodes at $St = 0.31$, $0.66$, $0.72$, and $0.77$ (panels a-d), plotting energy as a function of the dimensionless streamwise wavenumbers $k$ and the radial positions $r$. The white dashed lines represent dimensionless wavenumbers associated with the ambient speed of sound in the upstream and downstream directions. The KH wavepacket and guided jet mode are not periodic in the streamwise direction, instead exhibiting a spatial envelope that leads to energy spreading in wavenumber space. For frequencies near the screeching conditions ($St = 0.66, 0.72, 0.77$), the KH wavepacket manifests as a peak observed near the shear layer, with support extending into the core region, at positive wavenumbers in the range $k_{kh} \in [5, 8]$, corresponding to a phase speed of around $0.7 U_j$. In contrast, the guided jet mode has its highest amplitude in the core region, appearing at negative wavenumbers $k_{gj} \in [-6, -4]$, which corresponds to waves with a negative phase speed slightly slower than the ambient speed of sound $a_\infty$ \citep{nogueira2024guided}. On the other hand, the mode at $St = 0.31$ is confined primarily within the core region of the jet, lacking the shear-layer KH component typical of screech. The wavenumbers are also consistent with those expected for duct modes at this frequency, suggesting that the mode represents a resonance between duct and guided jet modes, as described by \citet{towne2017acoustic}.  We emphasize that while positive and negative phase speeds do not, in general, imply upstream- and downstream-traveling waves \citep{briggs1964, towne2015oneway}, they do for the wave families involved in screech \citep{Edgingtonetal21}, so we will use this terminology moving forward.

\subsection{Resolvent analysis} \label{sec:res}

Next, we compute resolvent modes for the screeching jet.  Since the global eigenmode representing the screech was found to be stable, it must be sustained by forcing from background turbulence.  Resolvent analysis identifies the optimal forcing and its response, yielding a harmonic screech mode that can be directly compared to data.  In particular, the harmonic modes obtained from resolvent analysis provide a better starting point for careful wavenumber analysis than the decaying global eigenmodes \citep{nichols2011global}.  

\begin{figure}
\centering
\includegraphics[scale=0.3]{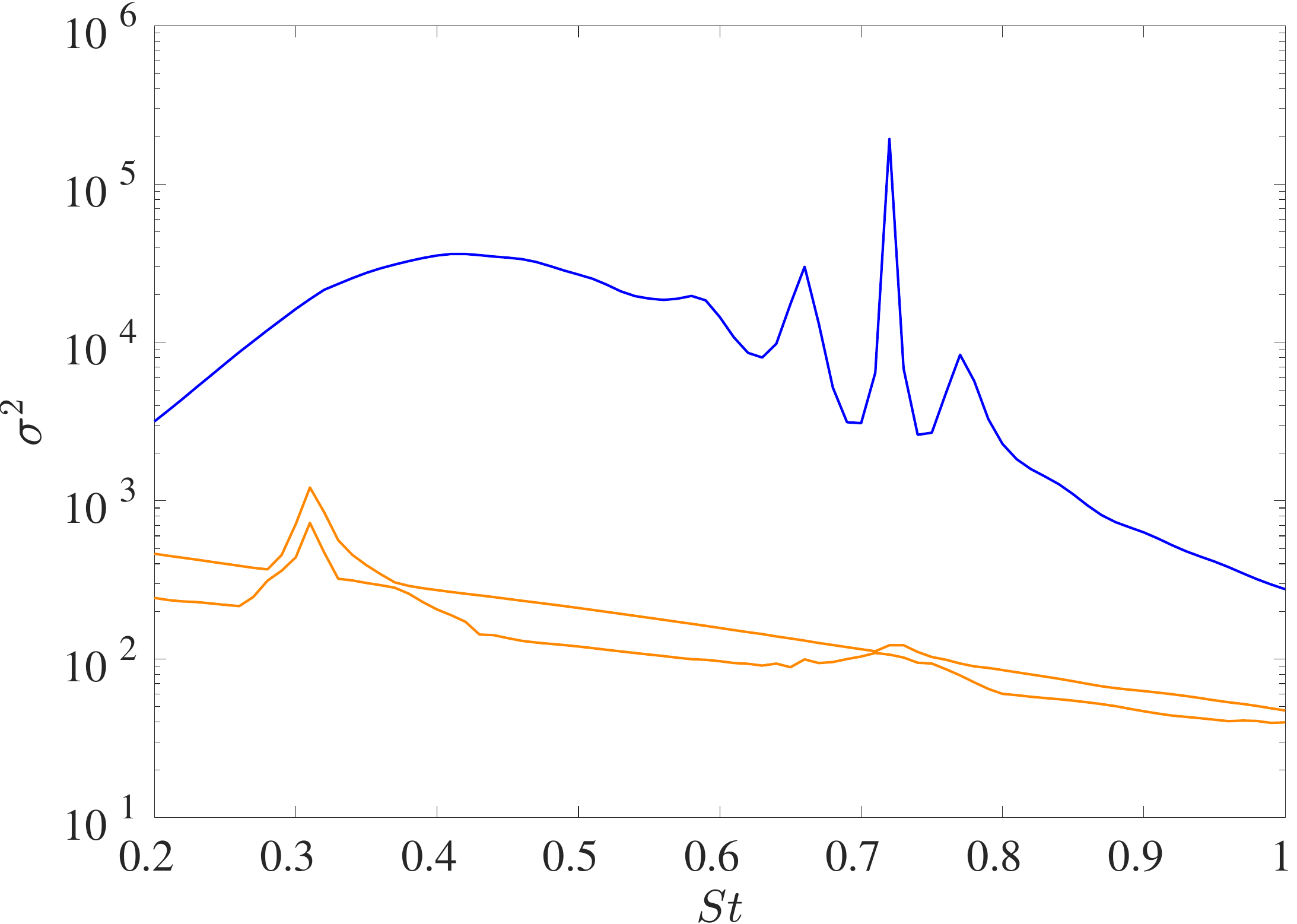}
\caption{The resolvent gain $\sigma^2$ of the screeching jet at $m=0$, obtained via linearization about the PIV mean flow: (blue) optimal gain $\sigma_1^2$; (orange) suboptimal gains $\sigma_2^2$ and $\sigma_3^2$.}
\label{fig:screechResolventGains}
\end{figure}

Resolvent modes are computed for the frequency range $ St \in [0.2, 1] $ with a resolution of $ \Delta St = 0.01 $ using $\bm A$ as the linearized Navier-Stokes operator from \S\ref{sec:setup}. Since the focus is on identifying the optimal modes, RSVD-LU is applied with $k = 5$ test vectors and $q=2$ power iterations for all frequencies. The resulting gain spectrum for $m = 0$ is shown in figure~\ref{fig:screechResolventGains}.  The leading gain matches trends observed in prior studies of subsonic jets \citep{Schmidtetal18, farghadan2024scalable}, while also revealing distinct peaks at $St = 0.66$, $0.72$, and $0.77$. These prominent peaks are a direct consequence of the interaction between the instability dynamics and the shock-cell structures within the flow, as indicated by the eigenspectrum analysis. The resolvent modes at these frequencies are associated with the screech phenomenon, which constitutes the central focus of this study.  The suboptimal gains are more than an order of magnitude lower for most frequencies, and their most prominent feature is a smaller peak at the duct-resonance frequency identified in the global eigenvalue analysis. The suboptimal resolvent response at $St = 0.31$ has a structure very similar to the global eigenmode at the same frequency (figure~\ref{fig:screechEigModes}), i.e.\ the duct-resonance mode; the peak appears in the suboptimal gain rather than the leading gain because the duct resonance is essentially uncorrelated with the broadband Kelvin-Helmholtz amplification at this frequency and therefore lives in a separate singular direction.  

\begin{figure} 
\centering 
\includegraphics[width=\textwidth]{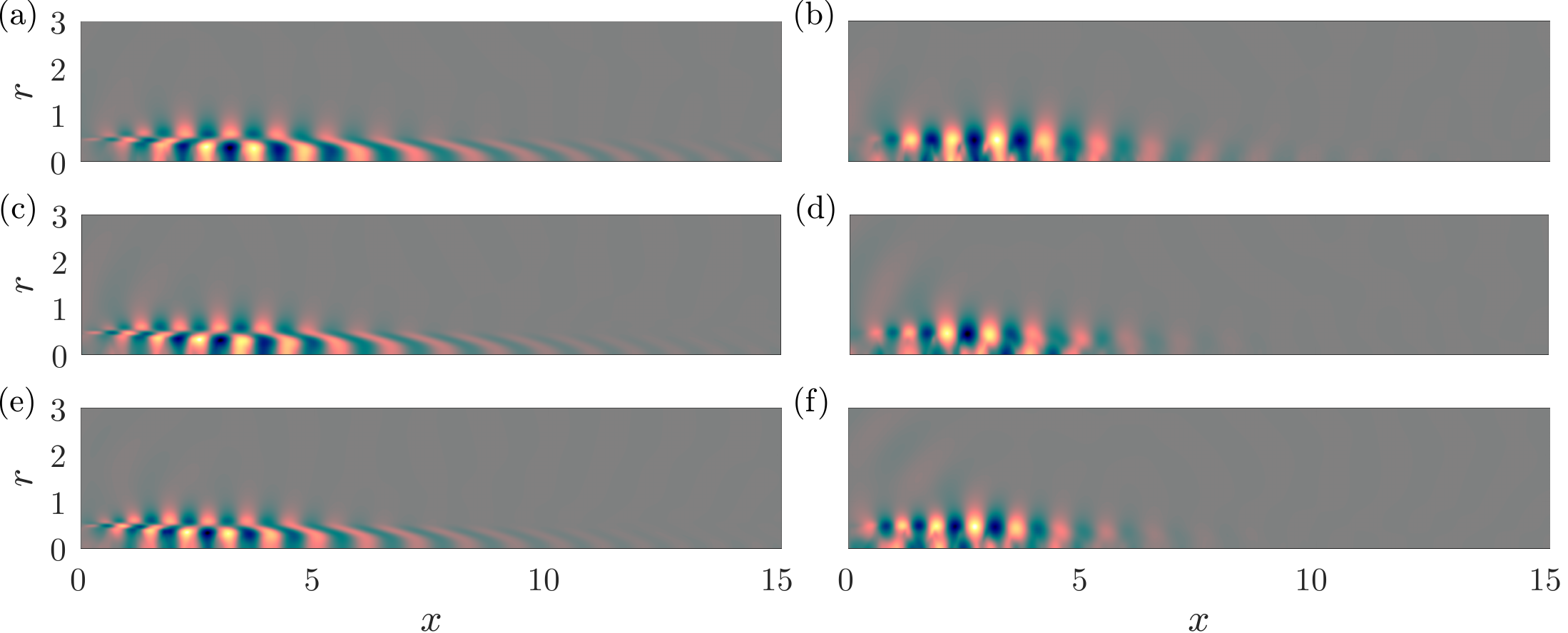} 
\caption{Resolvent response modes: (a, c, e) show the streamwise velocity, and panels (b, d, f) show the pressure component at frequencies $St = 0.66$, 0.72, and 0.77, respectively.} 
\label{fig:screechResolventModes} 
\end{figure}

\begin{figure}
  \centering
   \begin{overpic}[width=\textwidth]{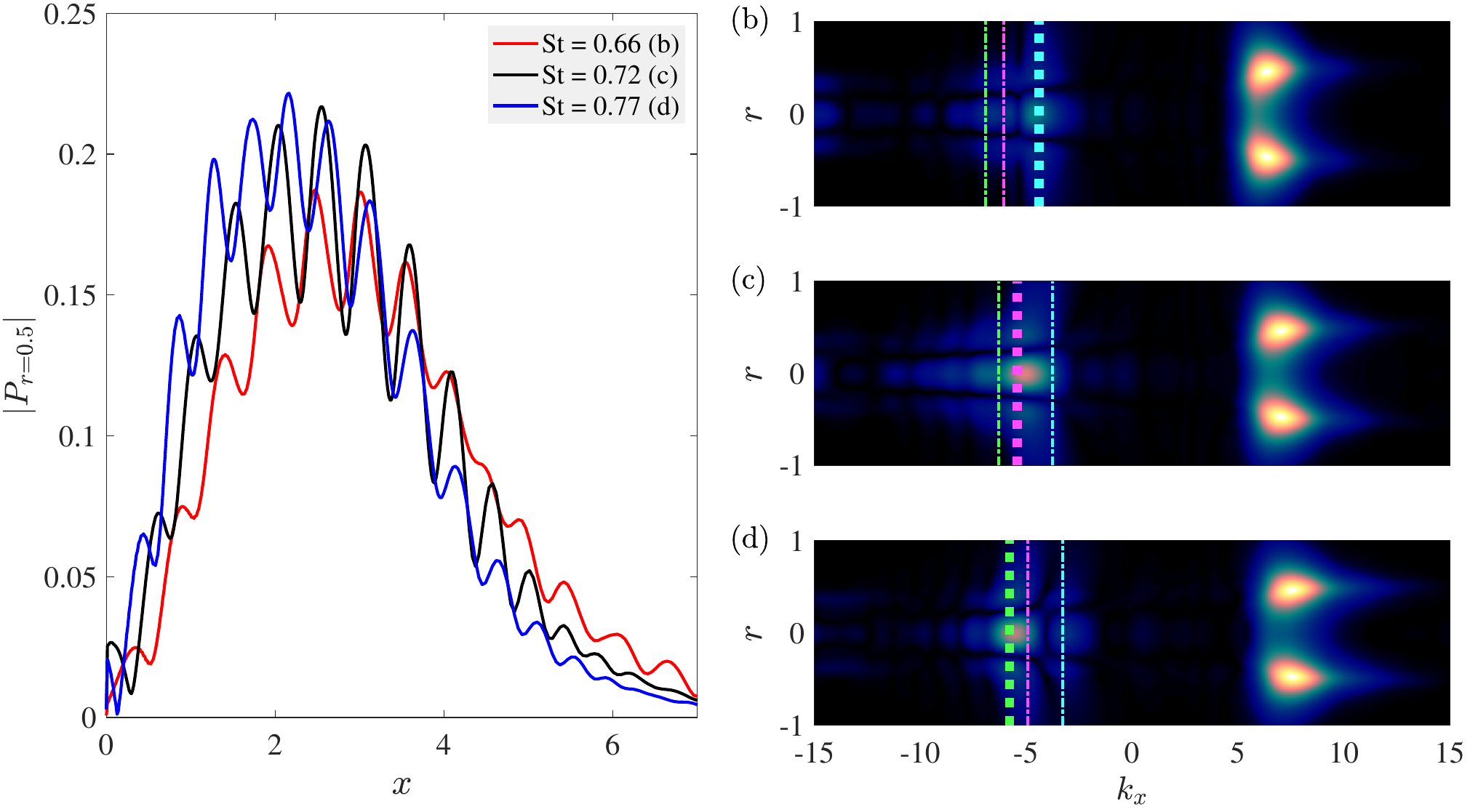}
    \put(-0.5,53.5){\small (a)}
   \end{overpic}
   \caption{Analysis of the resolvent response modes: (a) The absolute pressure along the lip line at $r/D = 0.5$ for $St = 0.66$ (red), $St = 0.72$ (black), and $St = 0.77$ (blue). Panels (b, c, d) show the wavenumber spectra of the optimal pressure responses at frequencies of 0.66, 0.72, and 0.77, respectively. Cyan, magenta, and green dashed lines correspond to $k_{kh} - k_{s_n}; n =1,2,3$.}
   \label{fig:screechResolventComp}
\end{figure}

The streamwise velocity and pressure components of the optimal resolvent responses at these frequencies are shown in figure \ref{fig:screechResolventModes}. As expected, these modes exhibit strong similarities to the eigenmodes computed at the same frequencies \citep{SchmidHenningson01}. A key distinction between the modes at the three frequencies lies in the location of the KH wavepacket envelope, which shifts upstream for higher $St$. This trend is evident in figure \ref{fig:screechResolventComp}(a), which plots the absolute value of the pressure along the lip line at $r = 0.5$. The maximum pressure amplitude occurs progressively closer to the nozzle as $St$ increases, with the peak for $St = 0.77$ closest to the nozzle, followed by $St = 0.72$, and $St = 0.66$ further downstream. This same trend is observed in the eigenmodes at the corresponding frequencies. In all three cases, the presence of a standing wave is observed, as expected given the superposition of upstream and downstream-propagating waves at each frequency. Additionally, figure \ref{fig:screechResolventComp}(b--d) displays the wavenumber spectra of the optimal responses, further illustrating the spatial distribution of these modes. The KH instability is clearly observed as a downstream-traveling wave concentrated near the shear layer, while the upstream-traveling guided jet mode is observed closer to the core region. The lowest frequency mode at $St = 0.66$ shows the smallest wavenumber separation between $k_s$ and $k_{gj}$, whereas the higher-frequency cases, $St = 0.72$ and $St = 0.77$, exhibit a larger wavenumber separation, suggesting interactions with smaller-scale shock cell structures located further downstream.

The shock-cell wavenumbers we use below, $k_s = [10.7, 12.3, 13.2]$ for the three leading resolvent modes at $St = [0.66, 0.72, 0.77]$, are obtained from a transform of the standing-wave pattern in the lip-line pressure (figure~\ref{fig:screechResolventComp}a), following \citet{Edgingtonetal22}; this method circumvents the resolution limit of a direct streamwise transform of the mean pressure (figure~\ref{fig:screechWaveNumShock}b), as we now describe.

As described by \citet{TamTanna82} and \citet{Edgingtonetal21, Edgingtonetal22}, the interaction between the streamwise wavenumbers of the KH instability, the shock cells, and the guided jet mode is governed by \begin{equation} k_{gj} = k_{kh} - k_s, \label{eqn:WaveNum2} \end{equation} where $k_{kh}$, $k_s$, and $k_{gj}$ represent the wavenumbers of the KH instability, shock cells, and guided jet mode, respectively. This equation shows that the downstream-traveling KH wave interacts with the shock cells, and the resulting wavenumber difference gives rise to the upstream-traveling guided jet mode. The rapid variation in shock-cell spacing, as demonstrated in figure \ref{fig:screechWaveNumShock}(a), which shows the mean streamwise velocity profile with weak shock cells and dashed lines indicating their spacing, means that rather than the shock structures being represented by a single value of $k_s$, they are represented by a broad primary peak and a series of suboptimal peaks. As shown in figure \ref{fig:screechWaveNumShock}(b), the wavenumber spectrum of the mean pressure along the centerline at $r = 0$ exhibits a dominant peak at $k_s \approx 11$, with cyan dashed lines marking local maxima. The computation of the shock cell wavenumber $k_s$, is based on the mean pressure distribution, with a wavenumber resolution of $\Delta k_s = \pi/25 \approx 0.126$. Due to the gradual reduction in the size of the shock cells along the streamwise direction, the wavelength is not constant but spans a range, leading to a broad peak in the spectrum. This indicates a range of $k_s$ values that can satisfy \eqref{eqn:WaveNum2}. For rapidly varying shock-cell structures, such as those found in jets screeching in the A1 or A2 mode, identifying the first sub-optimal peak in the spectrum can be difficult if the primary peak is sufficiently broad. To circumvent the resolution limit of the streamwise transform, we instead identify the appropriate wavenumber through a transform of the standing wave pattern in the lip line absolute pressure shown in figure \ref{fig:screechResolventComp}; a key result of \citet{Edgingtonetal22} is that the wavenumber of the standing wave always matches the wavenumber of the region of local periodicity that closes the screech loop. The peak wavenumbers of the standing wave for the frequencies $St = [0.66,0.72,0.77]$ are $k_s=[10.7,12.3,13.2]$, whereas the visible peaks of the mean pressure spectra in figure \ref{fig:screechWaveNumShock}(b) are $k_s = [11.0,13.1]$. Thus, what appears to be the first suboptimal peak in the shock spectra is actually the second suboptimal, with the first suboptimal obscured by the breadth of the primary peak.

Predicted wavenumbers from \eqref{eqn:WaveNum2} are plotted on figure \ref{fig:screechResolventComp}(b,c,d), with the cyan, magenta, and green lines corresponding to the wavenumbers $k_{kh} - k_{s_n}; n = 1,2,3$, respectively (the subscript $n$ indexes the primary, first suboptimal, and second suboptimal shock-cell wavenumbers in order). It is clear that each of the leading resolvent modes corresponds to a triadic interaction between the KH wavepacket and different regions of quasi-periodicity in the flow; the highest-gain mode at $St = 0.72$, which matches the peak observed in experimental data, results from an interaction between the KH wavepacket and $k_{s_2}$, while the others correspond to interactions between the primary shock peak and the second suboptimal. To our knowledge, this is the first global stability or resolvent analyses to identify modes associated with multiple shock-cell wavenumbers, and the evidence here rests on the standing-wave extraction described above rather than on a visually resolved three-peak structure in the mean shock spectrum.

\begin{figure}
\centering
\includegraphics[width=\textwidth]{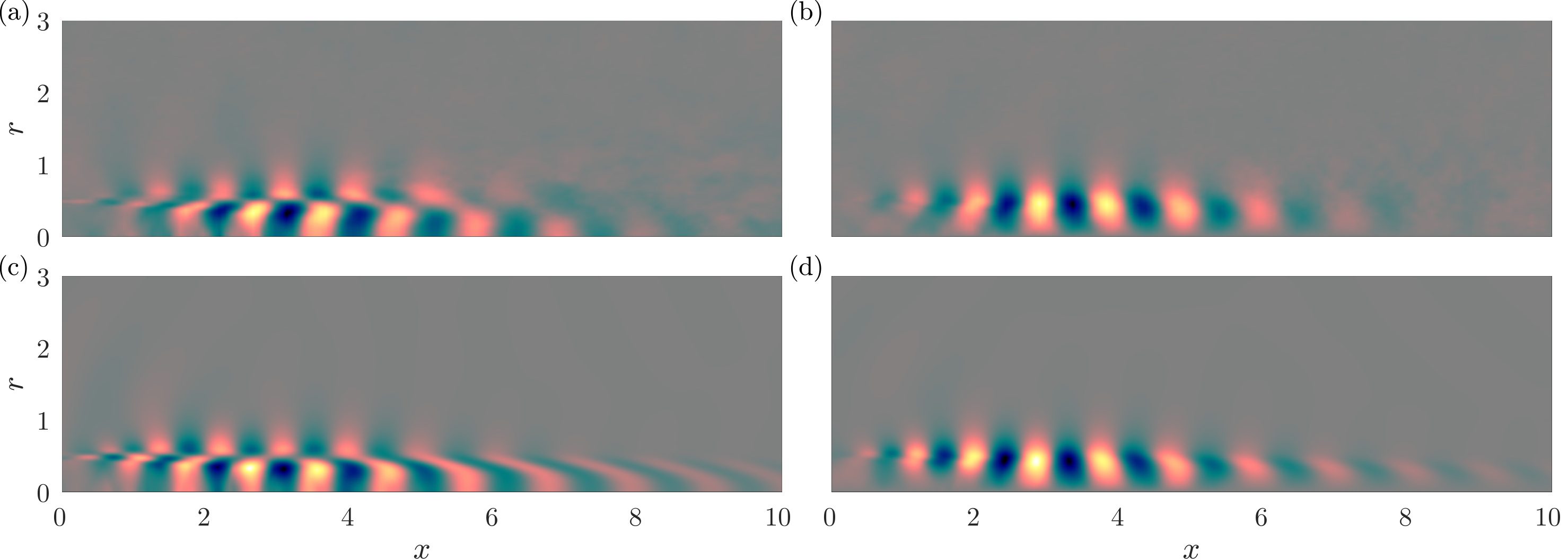}
\caption{Comparison between POD and resolvent modes at the screech frequency: (a, b) POD modes from experimental data and (c, d) optimal resolvent response modes at $St = St_s$. Panels (a, c) show the streamwise velocity components, and panels (b, d) display the radial velocity components.}
\label{fig:screechPODComp}
\end{figure}

In the next section, we will use the resolvent mode at the peak screech frequency as input for a harmonic resolvent analysis.  To further confirm that this mode accurately represents screech, we compare the velocity components of the computed resolvent mode and the leading POD mode extracted from experimental data, as shown in figure \ref{fig:screechPODComp}. The close agreement validates the numerical computations and substantiates the identification of the resolvent mode as the screech mode, which serves as the foundation for subsequent analyses. To complete the comparison at the screech frequency, we evaluate the normalized inner product between the eigenmode at $St = 0.72$ and the leading resolvent response in the Chu energy norm, $\langle \hat{\bm q}^{\rm eig}, \hat{\bm q}_1^R \rangle = 0.998$, and between the eigenmode and the leading POD mode, $\langle \hat{\bm q}^{\rm eig}, \hat{\bm q}^{\rm POD} \rangle = 0.89$. The first value confirms that the lightly damped eigenmode and the leading resolvent response are collinear at the screech frequency \citep{SchmidHenningson01}. The second is computed over the streamwise and radial velocity components on the PIV mesh, consistent with the convention used to set $A_s$, and matches the corresponding resolvent-POD inner product of $0.89$ obtained on the same protocol; the eigenmode and the leading resolvent therefore agree equally well with the POD-derived spatial pattern. Despite this near-collinearity, we use the resolvent mode rather than the eigenmode as the periodic base flow in the harmonic-resolvent analysis: the eigenmode has a nonzero damping rate and would need to be artificially made periodic for that purpose, whereas the resolvent response at $St_s$ is periodic by construction and is therefore the conceptually appropriate starting point.

\subsection{Harmonic resolvent analysis} \label{sec:harmonicres}

We showed in the previous section that resolvent analysis captures interactions between the shock cells, Kelvin-Helmholtz waves, and guided jet modes.  Next, we use the framework of harmonic resolvent analysis to study how the screech mode interacts with other fluctuations in the jet, and, in particular, how it leads to energy transfer to the mean flow and harmonics of the screech frequency, which cannot be captured using resolvent analysis. In what follows, ``standard resolvent mode'' refers to a mode of the resolvent operator defined in \S\ref{sec:resolvent}, ``harmonic resolvent mode'' refers to a mode of the harmonic-resolvent operator $\bm H$ defined in \S\ref{sec:harmonic}, and ``optimal'' designates the leading singular direction at a given frequency.

The screech mode oscillates at a well-defined frequency, and a modified mean flow $\bar{\bm q}_T$ is constructed by superimposing the screech mode $\hat{\bm q}_1^R$ obtained from resolvent analysis onto the original mean flow $\bar{\bm q}$. This is expressed as
\begin{equation}
\bar{\bm q}_T = \bar{\bm q} + \bm q_T,
\label{equation}
\end{equation}
where $\bm q_T = A_s \mathcal{R}(\hat{\bm q}_1^R e^{\text{i} 2\pi St_s t})$ represents the periodic component. Here, $St_s = 0.72$ is the screech frequency, $\mathcal{R}$ denotes the real part of the complex vector, and $A_s \approx 0.29$ is the relative amplitude of the screech mode with respect to the mean flow.  This relative amplitude was determined by matching the norm of the streamwise and radial velocity components of the resolvent modes to that of the leading POD mode extracted from experimental data shown in figure~\ref{fig:screechPODComp}(a,b). Because the POD modes derived from PIV contain only the streamwise and radial velocity components, the matching is restricted to those two components on the PIV mesh; the remaining state components (specific volume, azimuthal velocity, pressure) are not available from PIV and are not included. Using this modified mean flow, a time-periodic linearized Navier-Stokes operator $\bm A_T$ is constructed. Since the modified mean flow incorporates a single non-zero mode at the screech frequency and we use a bilinear formulation of harmonic resolvent analysis, the resulting $\bm{T}$ matrix contains only the non-zero components $\hat{A}_{p, 0}$ and $\hat{A}_{p, \pm 1}$.

The harmonic resolvent modes are computed using the RSVD-\Dt algorithm \citep{farghadan2024scalable} with $k=10$ test vectors and $q=1$ power iteration. These parameters enable accurate computation of the first few leading modes. For the time-stepping component, the RK4 integration scheme is used with a time step of $\Delta t = 0.011$, and a transient length of $T_t = 250$. We also employ the transient-removal strategy proposed by \citet{farghadan2024scalable}, which significantly reduces computational costs. Due to the real-valued nature of $\bm A_T$, the negative-frequency modes are the complex conjugates of the positive-frequency modes. As a result, only $\hat{A}_{p, 0}$ and $\hat{A}_{p, 1}$ are retained in memory, and the non-negative fluctuation frequencies are selected as $\Omega_f = \{0, St_s, 2St_s, 3St_s, 4St_s\}$. The response is computed up to the $3^{\text{rd}}$ harmonic $(St = 4St_s)$, with convergence observed by the $2^{\text{nd}}$ harmonic $(St = 3St_s)$, indicating that energy contributions from higher harmonics are negligible and do not influence the results. Hence, the analysis is confined to the frequency range within $\Omega = \Omega_q = \Omega_f$ for the remainder of the paper. 

The total CPU time for this analysis is less than 2 hours on 200 cores, using approximately 50 GB of memory. To place the memory requirements in context, the spatial state vector has $N = 5 \times N_x \times N_r = 5 \times 700 \times 300 = 1{,}050{,}000$ degrees of freedom per frequency, and the full block-structured harmonic-resolvent operator assembled over the nine frequencies $\{-4 St_s, -3 St_s, \ldots, 3 St_s, 4 St_s\}$ therefore has $9N \approx 9.45 \times 10^6$ rows. While RSVD-LU achieves comparable CPU efficiency for this problem, its memory requirements are prohibitively high. For the limited frequency range $ \Omega = \Omega_q = \Omega_f = \{0, St_s\} $, RSVD-LU already consumes around 3.5~TB of memory, making it impractical for broader frequency ranges. In contrast, RSVD-$\Delta t$ offers a more balanced approach, maintaining low memory usage without compromising computational efficiency, making it better suited for the wider frequency range analyzed in this study.

\begin{figure}
\centering
\includegraphics[width=\textwidth]{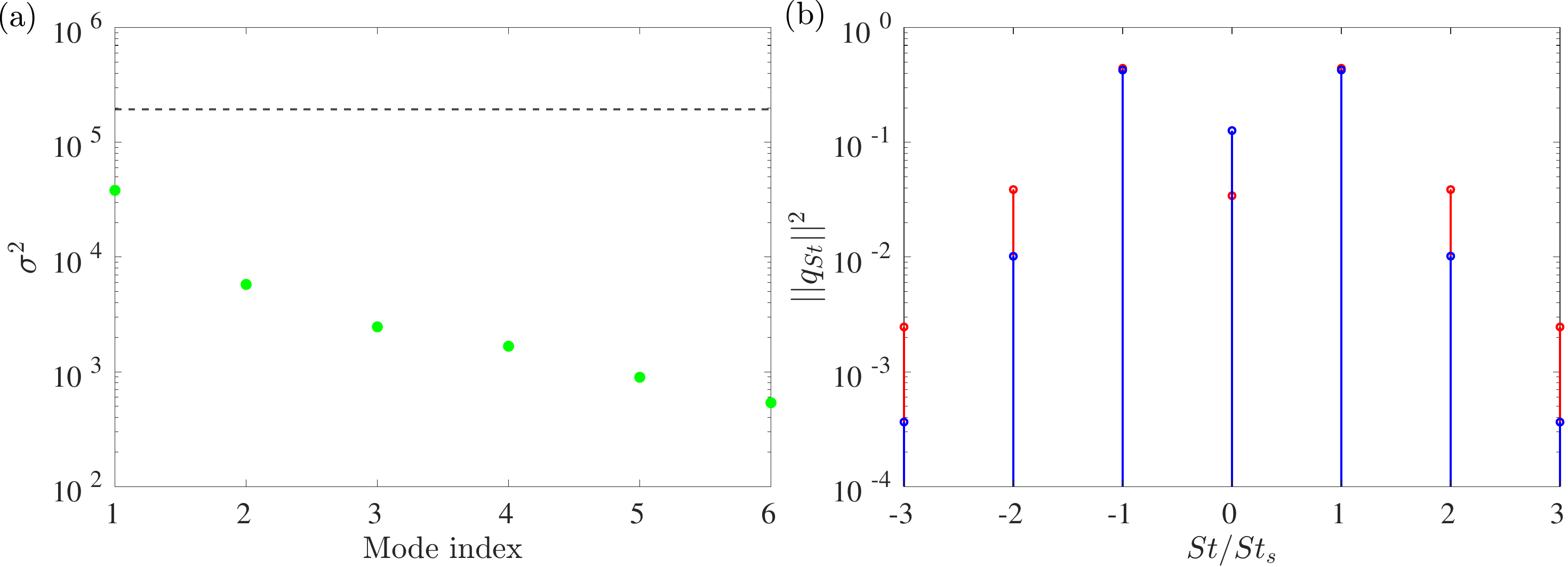}
\caption{Harmonic resolvent gains and energy distribution for the screeching jet: (a) The six leading harmonic-resolvent gains. For comparison, the horizontal dashed line shows the gain of the leading resolvent mode. (b) The energy spectrum of the optimal harmonic-resolvent response (red) and forcing (blue) modes. The $q$-norm of the response (with all frequencies combined) is 1.}
\label{fig:screechGainEnergy}
\end{figure}

In the following paragraphs, we report and discuss the results of the harmonic resolvent analysis.  In doing so, it is paramount to keep in mind the meaning of the harmonic resolvent modes: they represent the behavior of fluctuations (excluding the screech mode) whose dynamics are influenced by interactions with both the mean flow and the screech mode. Because the base flow $\bar{\bm q} + \bm q_T$ already carries content at $St = 0$ and $St = St_s$, and the harmonic-resolvent fluctuation $\bm q'$ represents motion not already in the base flow, the leading-mode response at these frequencies represents a modification to the assumed base flow. At $St = 0$, this reflects the modification of the mean flow induced by the screech mode interacting with itself. At $St = St_s$, the harmonic-resolvent response in general need not coincide with the screech mode (other coherent structures can live at the same frequency), and what we find is a structure that resembles the screech mode but with a meaningful phase difference.

Figure \ref{fig:screechGainEnergy}(a) shows the harmonic resolvent gains.  These are plotted as a function of mode index rather than frequency, since each mode may contain all of the frequencies retained in the analysis.  The gains exhibit two notable features.  First, the leading mode is more than an order of magnitude larger than the second mode; this gain separation suggests that the leading mode is likely to dominate the fluctuation (relative to the mean and the screech mode) for an arbitrary, unknown forcing.  Second, despite this gain separation, the leading harmonic resolvent gain is almost an order of magnitude smaller than the leading resolvent gain at the screech frequency (compare with the peak in figure~\ref{fig:screechResolventGains}).  This is consistent with the fact that the screech mode is by far the largest fluctuation (relative to the mean) observed in the experimental data.  

The energy distribution of the optimal harmonic-resolvent mode as a function of frequency, shown in figure~\ref{fig:screechGainEnergy}(b), highlights the role of the screech mode in distributing energy across frequencies. Unlike the optimal resolvent mode, which oscillates strictly at the screech frequency, the harmonic resolvent mode contains significant energy content at the screech frequency, while also feeding energy into the zeroth and higher harmonics through triadic interactions with the screech mode. In particular, the zeroth and first harmonics contain nearly half of the total energy of the mode. This observation underscores the pivotal role of the screech mode in redistributing energy across the frequency spectrum. 

\begin{figure}
\centering
\includegraphics[width=\textwidth]{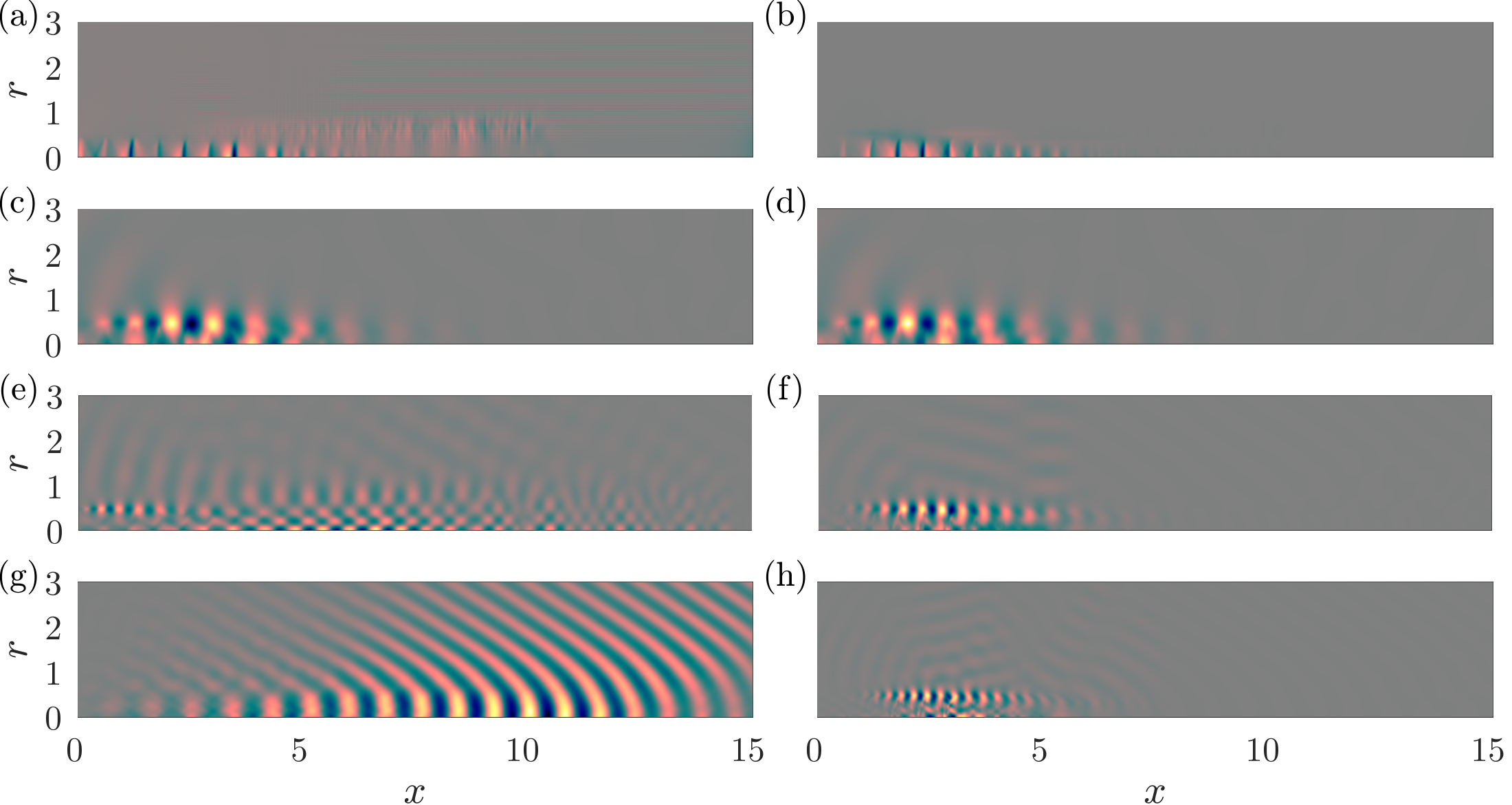}
\caption{Comparison between optimal resolvent and harmonic-resolvent response modes: (a, c, e, g) resolvent pressure modes and (b, d, f, h) harmonic-resolvent pressure modes at frequencies of $St = 0, St_s, 2St_s$, and $3St_s$, respectively.}
\label{fig:screechModes}
\end{figure}

Next, we analyze the spatial structures of the leading harmonic resolvent mode at each frequency and compare them with those of the standard resolvent modes at the same frequencies to understand their physical characteristics. Recall that, due to the large gain separation in the harmonic resolvent gain spectrum, the leading harmonic resolvent mode shapes represent the dominant physical structures that we expect to observe in the fluctuation field (relative to the mean flow and screech mode).  The standard resolvent modes were computed using the RVSD-LU algorithm with $k=5$ test vectors and $q=2$ power iterations. Results are shown in figure~\ref{fig:screechModes} and are discussed in order of decreasing energy content within the harmonic resolvent mode.

At the screech frequency, $St = St_s$, the resolvent mode features the signatures of a downstream-traveling KH wavepacket and an upstream-traveling guided jet mode, as discussed in detail in \S\ref{sec:res}.  The harmonic resolvent mode is visually similar, exhibiting the same two waves. A quantitative comparison confirms this close agreement; projecting the two modes onto one another yields a normalized inner product of 0.97, indicating that they are nearly collinear in the chosen norm. Together, these results indicate that the easiest structure to force within the screeching jet is a modification to the screech mode itself.  Physically, this fluctuation is enabled by the same resonance loop underlying the screech mode via its interaction with the mean flow, but is at the same time modulated by the presence of the screech mode, leading to energy redistribution to other frequencies and a lower gain than the screech mode itself.  

At zero frequency, $St=0$, the resolvent mode contains modifications to the shock cells throughout the potential core as well as likely unphysical high-wavenumber noise within the downstream turbulent region of the jet.  The gain of this mode is low, making it both difficult to converge numerically and unlikely to play a prominent role in flow physics.  In contrast, the harmonic resolvent mode consists entirely of a modification to the shock cells focused in the region where the screech mode is active, and the amplitude of this component of the mode is significant.  

At the first harmonic, $St = 2 St_s$, the resolvent mode consists of a KH wavepacket near the nozzle (consistent with the higher frequency) and guided jet modes with higher radial order \citep{towne2017acoustic}.  The harmonic resolvent mode is completely different, consisting of a wavepacket in the region where the screech mode is active and acoustic waves propagating perpendicular to the jet axis.  These acoustic waves have been observed in experiments \citep{powell1953mechanism} but have not been previously captured within linearized Navier-Stokes analyses.  The present results show that the interaction of fluctuations with the screech mode can generate the observed acoustic signature.  At the second harmonic, $St = 3 St_s$, the resolvent mode consists of direct acoustic radiation.  Again, the harmonic resolvent mode is completely different, consisting of a concentrated waepacket in the region where the screech mode is active and a beam of acoustic radiation that is again consistent with experimental observations \citep{tam2014harmonics}.  Overall, the differences between resolvent and harmonic resolvent modes indicate the consequential impact of the screech mode on the dynamics of other fluctuations in the jet, including the redistribution of energy to other frequencies.

\begin{figure}
\centering
\includegraphics[width=\textwidth]{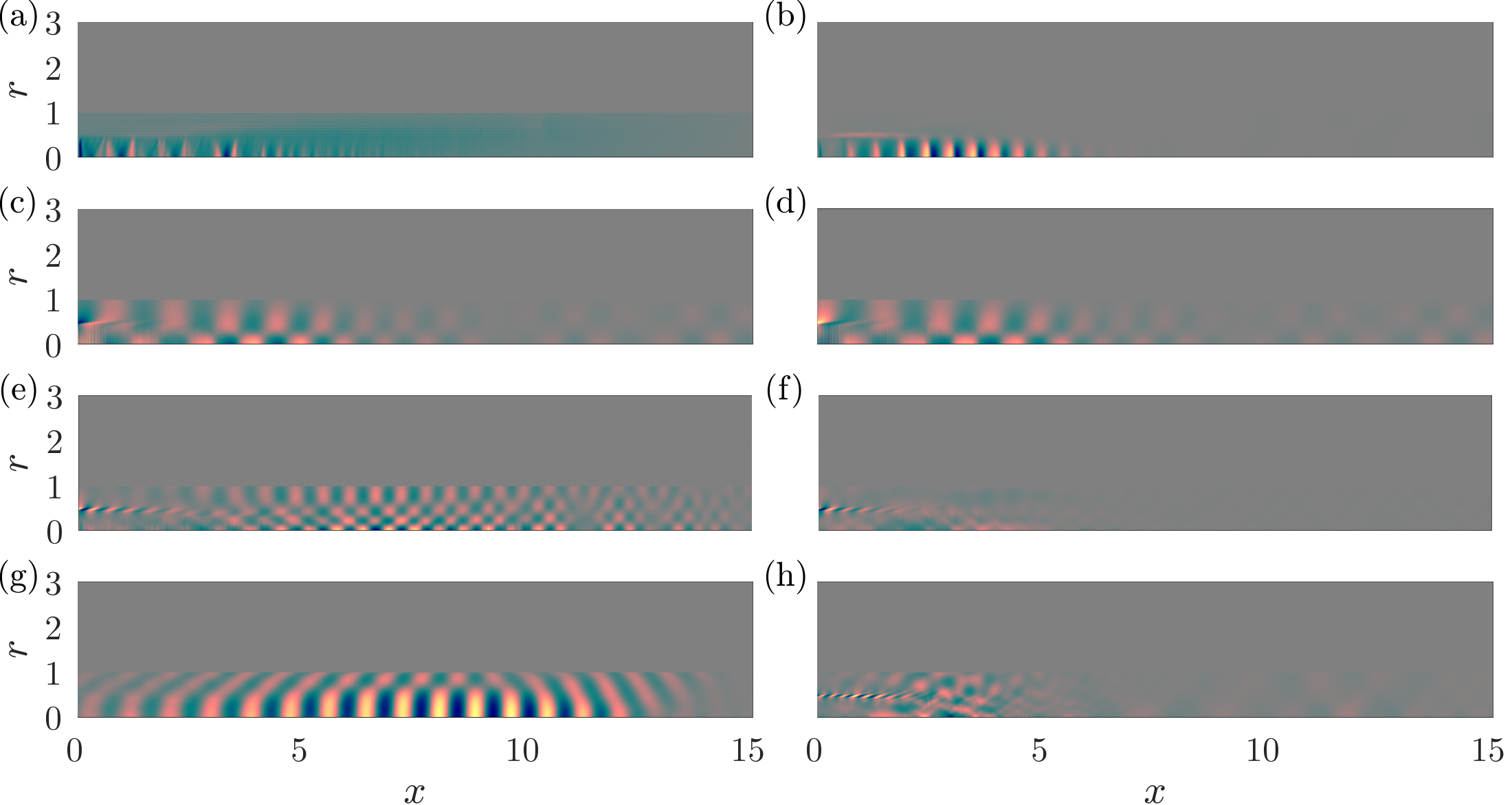}
\caption{Comparison between optimal resolvent and harmonic-resolvent forcing modes: (a, c, e, g) resolvent pressure-forcing modes and (b, d, f, h) harmonic pressure-forcing modes at frequencies of $St = 0, St_s, 2St_s$, and $3St_s$, respectively.}
\label{fig:screechForcing}
\end{figure}

Figure \ref{fig:screechForcing} compares the optimal forcing modes from resolvent and harmonic resolvent analyses. At the screech frequency, $St = St_s$, the forcing modes are nearly identical, consisting of slanted structures in the near-nozzle region that effectively excite the KH waves \citep{Schmidtetal18} and more distributed acoustic-like structures that effectively excite the guided jet mode due to the self-adjoint nature of acoustic waves.  The similarity between the resolvent and harmonic resolvent forcing mode reiterates the point that the best way to generate fluctuations at the screech frequency is to excite the screech resonance, even with the redistribution of energy due to the presence of the screech mode itself.  

The forcing modes at zero frequency, $St=0$, are consistent with the response modes at the same frequency, with the former spread throughout the potential core and the latter concentrated in the region where the screech mode is active.  At both of the harmonic frequencies, $St = 2 St_s$ and $3 St_s$, the resolvent forcing modes are dominated by the direct forcing of acoustic waves.  In contrast, the tilted structures that effectively force the KH waves are more prominent in the harmonic resolvent forcing modes.  Given the low gains of the resolvent modes at the non-screech frequencies and the differences in the forcing modes, we conclude that the harmonic resolvent response modes at non-screech frequencies are primarily driven by energy transfer due to the presence of the screech mode, rather than direct linear amplification.

\begin{figure}
\centering
\includegraphics[width=\textwidth]{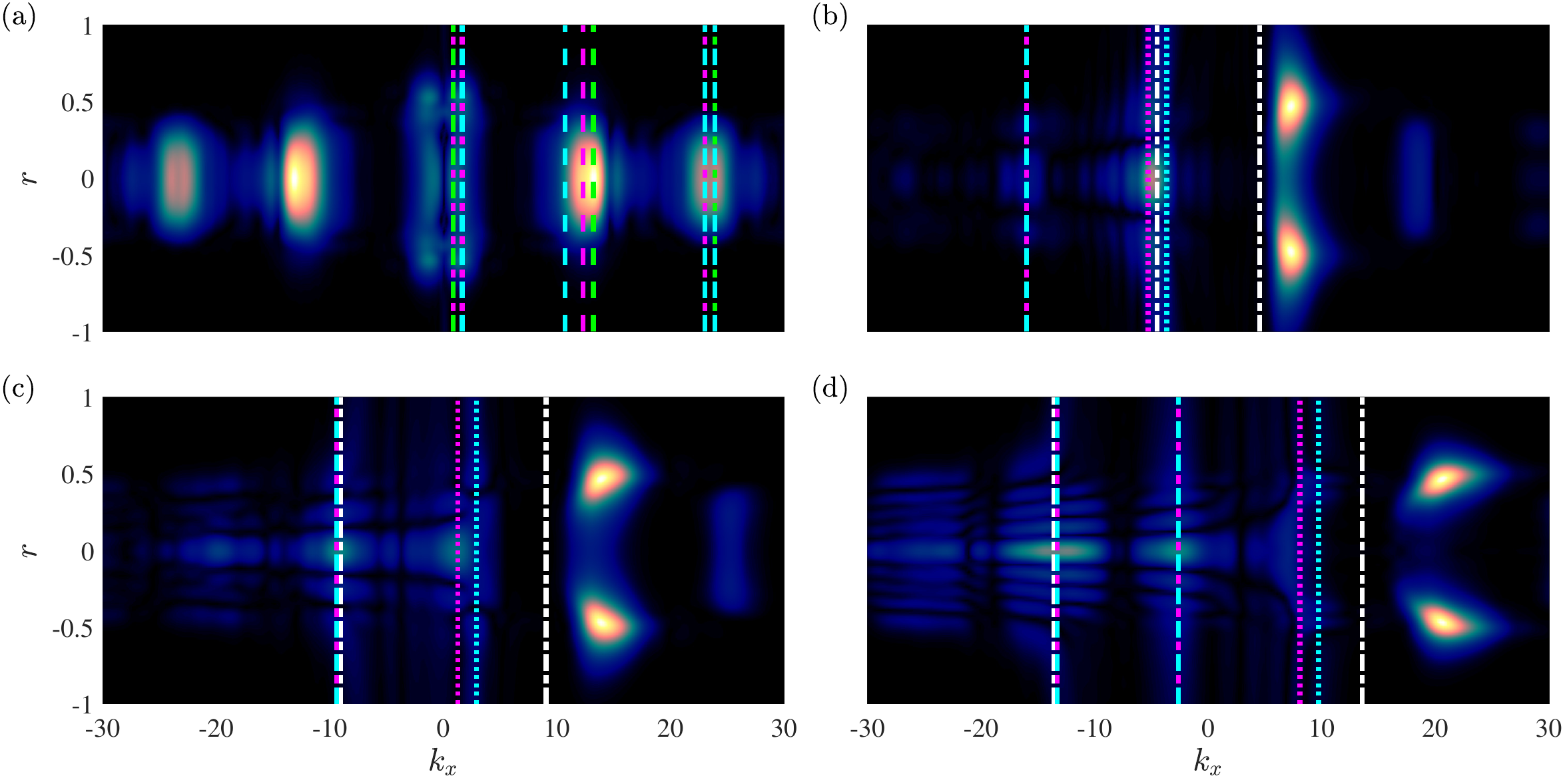}
\caption{The wavenumber spectra of the pressure component of the optimal harmonic-resolvent response at (a-d) the frequencies $St = 0, St_s, 2St_s$, and $3St_s$, respectively. The vertical dashed lines indicate wavenumbers for (white) acoustic waves and (cyan, magenta, green) $k_{kh} - k_{s_n}$; $n=1, 2, 3$. }
\label{fig:screechWaveNumHarmonicModes}
\end{figure}

Finally, the wavenumber spectra of the optimal harmonic resolvent response modes, shown in figure \ref{fig:screechWaveNumHarmonicModes}, provide insights into the wave interactions responsible for these modes. At the screech frequency, $St = St_s$, the wavenumber spectrum is qualitatively similar to that of the resolvent screech mode (shown previously in figure~\ref{fig:screechResolventComp}); there are minor differences in the peak wavenumber of the KH wavepacket, but little else distinguishes the two modes. Weaker peaks appear at the negative wavenumber $k_{kh}-(k_{s_{1}}+k_{s_{2}})$ and the positive wavenumber $k_{kh} + k_{s_{1}}$ that are confined within the core of the jet, consistent with downstream-traveling duct modes \citep{towne2017acoustic}.


At $St=0$, the harmonic resolvent mode is real-valued and mirror-symmetric about zero, representing the stationary modification of the mean flow and shock cells by the unsteady structures active at the screech frequency. The wavenumber spectrum displays dominant peaks at shock-associated wavenumbers (\eg $k_{s_{2}}$ and $k_{s_{3}}$) as well as high-amplitude peaks at various sum and difference combinations. While the individual peaks suggest complex interactions, the spectrum exhibits a consistent spacing of approximately $\Delta k = 11$, beginning at $k_{mod} = 15$. This starting wavenumber corresponds to the self-interaction of the Kelvin-Helmholtz wavepacket observed at the screech frequency ($2k_{kh}$, where $k_{kh} \approx 7.5$). This spectral signature confirms a two-way coupling mechanism: the shock structures modulate the KH wavepacket \citep{nogueira2022wave}, and the non-linear self-interaction of this modulated wavepacket in turn modifies the mean flow.


At the first harmonic, $St = 2St_s$, in addition to the KH wavepacket appearing at twice the wavenumber of the fundamental, there are peaks associated with both $k_{kh}-k_{s_{1}}$ and $k_{kh}-k_{s_{2}}$; the former has strong support outside of the jet, whereas the latter has the highest amplitude in the jet core. These wavenumbers, close to zero, are consistent with the directivity of the beams of acoustic radiation observed in figure \ref{fig:screechModes}(f) and in \citet{tam2014harmonics}. Interestingly, the results here suggest that while $k_{s_{2}}$ is responsible for closing the resonance loop, much of the sideline radiation of the harmonic is in fact associated with an interaction between $k_{kh}$ and $k_{s_{1}}$. There is also a component of upstream radiation associated with $k_{kh}-(k_{s_{1}}+k_{s_{2}})$, which we again suggest is the signature of a shock-modulated wavepacket undergoing triadic interaction with another shock wavenumber.

At the second harmonic, $St = 3St_s$, the KH wavepacket peaks at  $k_x \approx 21$, approximately triple the value observed at $St = St_s$, as expected for a second harmonic. Here, the peaks associated with the interactions $k_{kh}-k_{s_{1}}$ and $k_{kh}-k_{s_{2}}$ are both relatively weak; there is some redistribution of energy to these wavenumbers, but no clear local peak. The stronger peaks in fact occur at $k_{kh}-(k_{s_{1}}+k_{s_{2}})$ and $k_{kh}-(2k_{s_{1}}+k_{s_{2}})$. We suggest that these results can be interpreted as the KH wavepacket, modulated by its interaction with $k_{s_{2}}$, interacting with $k_{s_{1}}$ and its harmonic, respectively.

\subsection{Nonlinear forcing} \label{sec:nonlinRes}

The harmonic resolvent analysis in the previous section examined the response of the jet to the optimal forcing, which is the sensible approach when the true nonlinear forcing $\hat{\bm f}$ is unknown. However, we showed in \S\ref{sec:nonlinear} that for a bilinear system, the forcing is not entirely unknown.  Specifically, our bilinear formulation allows us to directly study how the nonlinear self-interaction of the screech mode, $\hat{\bm f}_{nl} = \boldsymbol{\mathcal{B}}(\bm q_T, \bm q_T)$, impacts the other fluctuations in the jet. This approach helps elucidate the role of the screech mode in redistributing energy to harmonic frequencies. Throughout this section, the action of $\bm H$ on $\hat{\bm f}_{nl}$ is computed using time-stepping analogous to the RSVD-\Dt algorithm.

\begin{figure}
\centering
\includegraphics[width=\textwidth]{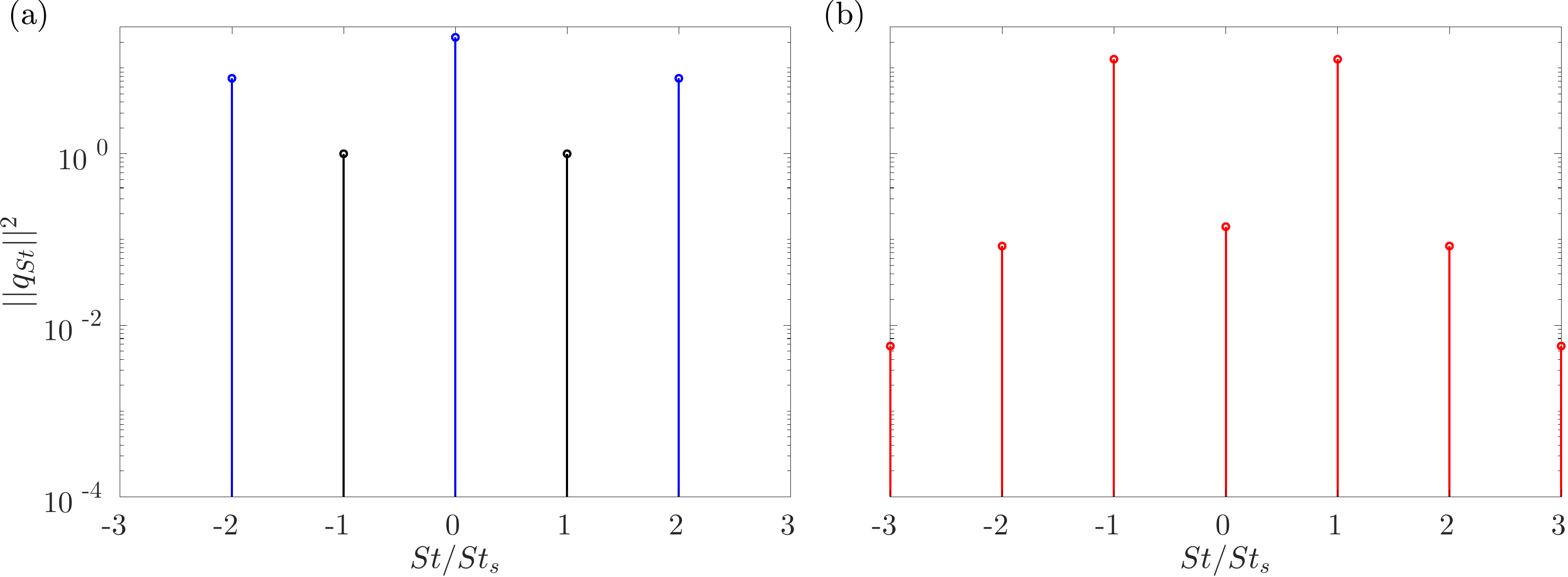}
\caption{Energy distribution of quantities in the nonlinear analysis: (a) $\hat{\bm q}_1^R$ (black), $\hat{\bm f}_{nl}$ (blue), (b) $\hat{\bm q}_{nl}$ (red).}
\label{fig:screechNonlinearForcingNorms}
\end{figure}

According to \eqref{eqn:nonlinearF}, the forcing term $\hat{\bm f}_{nl}$ is nonzero at $St = 0$ and $St = \pm 2St_s$ due to the concentration of $\hat{\bm q}_1^R$ at $St = \pm St_s$ and the nonzero nature of $\hat{\bm A}_i$ for $|i| \leq 1$. Figure \ref{fig:screechNonlinearForcingNorms}(a) illustrates this behavior by showing the energy spectrum of both $\hat{\bm q}_1^R$ and $\hat{\bm f}_{nl}$ across different frequencies. The normalization $||\hat{\bm q}_{\pm 1}||^2 = 1$ ensures that the total energy of $\hat{\bm q}_1^R$ is $||\hat{\bm q}_1^R||^2 = 2$. More generally, the total energy of any state $\bm q$ across all frequencies is defined as  
\begin{equation}
|| \hat{\bm q} || ^ 2 = \sum_{i = -3}^{3} || \hat{\bm q}_i ||^2,
\label{eqn:total_energy}
\end{equation}  
where $\hat{\bm q}_i$ represents the state at the $i^{\text{th}}$ harmonic of the screech frequency. The total energy of the nonlinear forcing term is computed $||\hat{\bm f}_{nl}||^2 = 38.0$. Figure \ref{fig:screechNonlinearForcingNorms}(b) presents the energy distribution of the nonlinear response $\hat{\bm q}_{nl}$, revealing that its total energy is $||\hat{\bm q}_{nl}||^2 = 25.6$, which is lower than the input forcing norm. This indicates that the forcing undergoes some degree of attenuation.  Notably, while the nonlinear forcing is zero at $St = St_1$ and $St_3$, the triadic interactions retained within harmonic resolvent analysis redistribute energy to these frequencies.

\begin{figure}
\centering
\includegraphics[width=\textwidth]{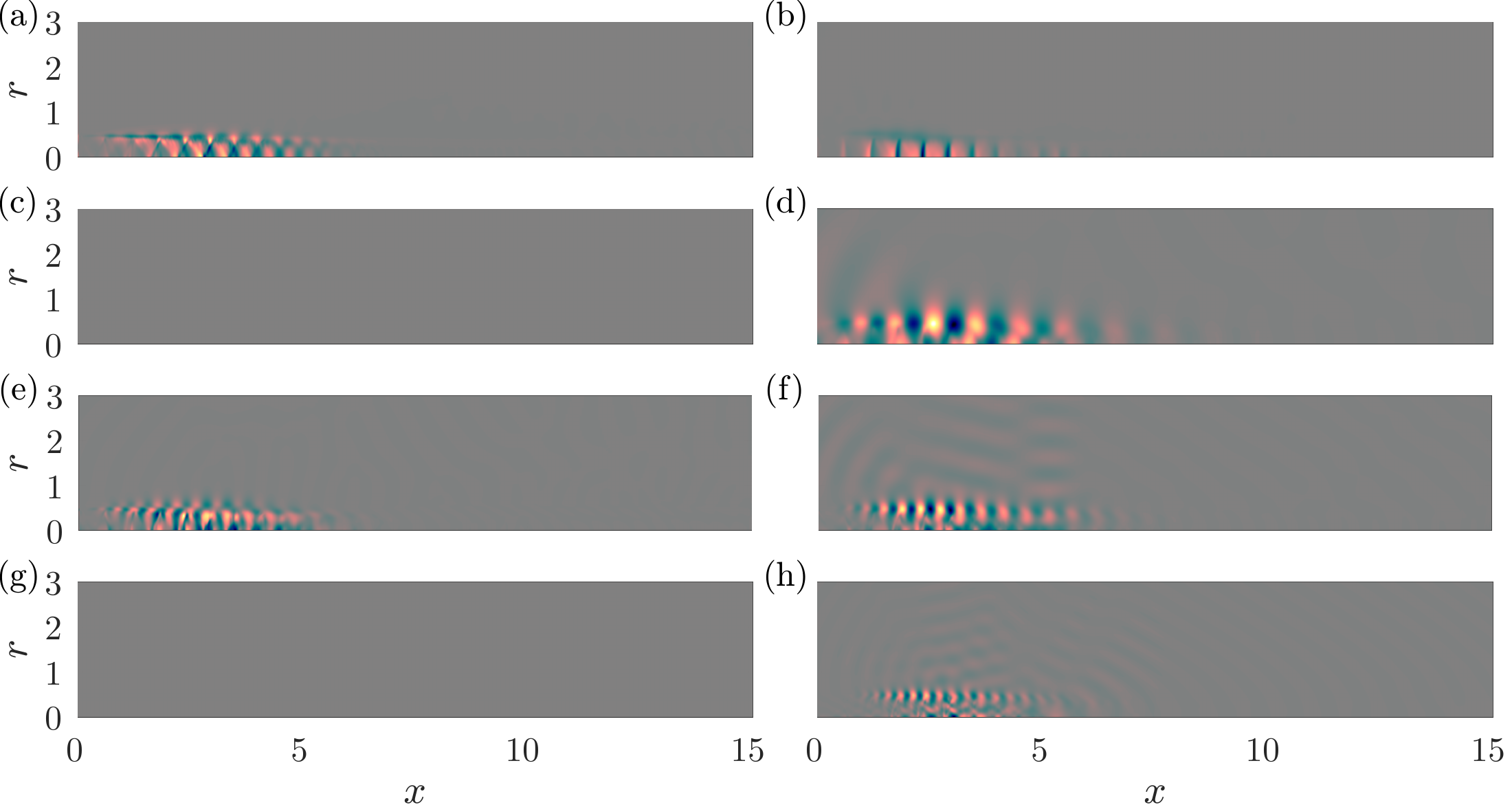}
\caption{Spatial structure of quantities in the nonlinear analysis: (a, c, e, g) forcing pressure modes and (b, d, f, h) corresponding response pressure modes at frequencies $St = 0, St_s, 2St_s$, and $3St_s$, respectively.}
\label{fig:screechNonlinearForcing}
\end{figure}

The pressure modes of both $\hat{\bm f}_{nl}$ and $\hat{\bm q}_{nl}$ are depicted in figure \ref{fig:screechNonlinearForcing}. While the nonlinear forcing modes at $St = 0$ and $St = \pm 2St_s$ differ significantly from the optimal harmonic forcing modes in figure \ref{fig:screechForcing}, the pressure response modes of $\hat{\bm q}_1^H$ (figure \ref{fig:screechModes}) and $\hat{\bm q}_{nl}$ closely resemble each other.  To quantify this agreement, we compute the inner products between $\hat{\bm q}_{1, i}^H$ and $\hat{\bm q}_{nl, i}$ for $|i| \leq 3$, obtaining values greater than 0.95 in all cases, indicating a strong alignment at each individual frequency. However, the inner product $\langle \hat{\bm q}_1^H, \hat{\bm q}_{nl} \rangle = 0.36$ suggests a phase mismatch at some frequencies. Here, the values greater than $0.95$ are normalized inner products $\langle \hat{\bm q}_{1,i}^H, \hat{\bm q}_{nl,i} \rangle$ evaluated frequency by frequency, while the value $0.36$ is the normalized inner product of the full frequency-concatenated modes. The two differ because the harmonic-resolvent SVD selects the relative phase between frequency blocks that maximizes the total response, whereas $\hat{\bm q}_{nl}$ inherits whatever phase relationship the known nonlinear self-interaction produces; this phase difference is meaningful, not arbitrary. The two analyses are therefore complementary: the harmonic-resolvent SVD identifies the optimal receptive direction $\hat{\bm q}_1^H$ independently of the forcing, while the present analysis reveals the response to the specific known nonlinear self-interaction, which is best interpreted as a correction to that optimal direction rather than as exciting it directly.    

To summarize, this analysis shows that the forcing provided by the direct nonlinear self-interaction of the screech mode, along with its triadic interactions with other frequencies, is sufficient to explain the redistribution of energy to other frequencies and the associated properties of these fluctuations.  These include modifications to the shock cells and screech mode, and the experimentally observed acoustic beams at the harmonic frequencies.  No appeal to an unknown forcing by background turbulence is necessary to explain or model these phenomena.  

\section{Conclusions}
\label{sec:conclusion}

This investigation into the screech phenomenon of a supersonic jet leverages a combination of global stability analysis, resolvent analysis, harmonic resolvent analysis, and a nonlinear extension of harmonic resolvent analysis, complemented by experimental PIV data, to elucidate the underlying dynamics of jet screech. Our findings provide a detailed understanding of the feedback mechanisms, wave interactions, and energy redistribution processes that define this aeroacoustic phenomenon, offering new insights into its complexity and implications for noise mitigation.

First, we used global stability analysis of the linearized Navier-Stokes operator to identify the least stable eigenmode, which appeared at a Strouhal number of $St_s = 0.72$, close to the experimentally observed screech frequency of $St = 0.714$. This mode, characterized by a downstream-propagating KH wavepacket and an upstream-traveling guided jet mode, encapsulates the core feedback loop driving screech. Beyond the primary mode, additional eigenmodes were identified at $St = 0.66$ and $St = 0.77$, suggesting that multiple resonance loops can coexist within the jet, each associated with distinct interactions between KH instabilities and shock-cell structures. However, the dominance of the $St = 0.72$ mode corresponds to the single, prominent screech tone observed experimentally, highlighting its role as the most energetically significant instability.

Second, we used resolvent analysis to further illuminate the amplification characteristics of the jet's dominant structures. The gain spectrum exhibited pronounced peaks near the screech frequency, with optimal response modes at $St = 0.66, 0.72$, and $0.77$ showing strong agreement with POD modes derived from PIV data. These modes, localized primarily in the shear layer, feature standing-wave patterns indicative of the superposition of upstream- and downstream-traveling waves. Wavenumber analysis confirmed that the screech mechanism arises from triadic interactions in which KH wavepacket at wavenumber $k_{kh}$ interacts with shock-cell wavenumbers $k_s$ to produce the guided jet mode wavenumber $k_{gj}$, as governed by $k_{gj} = k_{kh} - k_s$.

Third, we employed harmonic resolvent analysis, integrating the screech mode obtained from resolvent analysis into a time-periodic base flow to explore cross-frequency triadic interactions. This approach revealed a significant redistribution of energy, with the screech frequency ($St_s = 0.72$) transferring energy to the zeroth, first ($2 St_s$), and second ($3 St_s$) harmonics via triadic interactions. A striking outcome was the identification of localized acoustic waves at $2 St_s$ and $3 St_s $, radiating at oblique angles from the jet axis, features consistent with experimental observations but previously lacking a mechanistic explanation. These acoustic beams stem from interactions between the KH wavepacket and multiple shock-cell wavenumbers, with the second harmonic's sideline radiation linked to $k_{kh} - k_{s_1}$ and the third harmonic showing contributions from more complex combinations like $k_{kh} - (k_{s_1} + k_{s_2})$.

Fourth, we used a novel extension of harmonic resolvent analysis to demonstrate the central role of the screech mode in energy redistribution. By introducing the nonlinear forcing term $\hat{\boldsymbol{f}}_{nl} = \mathcal{B}(\boldsymbol{q}_T, \boldsymbol{q}_T)$, which arises naturally within a bilinear formulation, into the harmonic resolvent framework, we found that the resulting response modes closely matched the optimal harmonic resolvent modes, despite the differences in forcing distribution. This shows that the nonlinear self-interactions of the screech mode, along with its triadic interactions with other frequencies, govern the dominant response structures across the frequency spectrum and are sufficient to explain all of the observed phenomena, reinforcing its role as the primary driver of energy transfer in the system.

Each of the three variations of resolvent analysis can be understood in terms of the interactions it captures.  Resolvent analysis captures triadic interactions between the shock cells and the Kelvin-Helmholtz wave and guided jet mode at the screech frequency, which represent the essential components of the screech resonance loop.  Harmonic resolvent analysis additionally captures triadic interactions between the screech mode (consisting of the Kelvin-Helmholtz wave and guided jet mode) and additional fluctuations at the screech frequency, zero frequency (mean-flow correction), and harmonics of the screech frequency, enabling inter-modal energy transfer.  Finally, our nonlinear forcing analysis captures the impact of the screech mode's nonlinear self-interaction on these other fluctuations in the jet.  

On the computational front, this study employs RSVD-\Dt \citep{farghadan2024scalable, farghadan2024scalable} for both harmonic resolvent analysis (which takes only a few hours) and for computing the action of $\bm H$ on the nonlinear forcing (which takes only several minutes). This approach avoids the memory-intensive LU decomposition typically required by traditional methods. It enables efficient computations across multiple harmonics, with convergence achieved at the second harmonic, indicating that higher harmonics contribute negligibly in this configuration. The scalability of RSVD-\Dt makes it a powerful tool for future investigations of complex, periodic flows where capturing triadic interactions is critical.

In summary, our results demonstrate that while linear tools such as resolvent analysis capture the primary screech instability, a full understanding of the phenomenon requires accounting for triadic interactions via harmonic resolvent analysis. These interactions govern energy transfer, suppress harmonic amplification, and generate the characteristic acoustic radiation patterns observed in screeching jets. Moreover, our bilinear formulation of harmonic resolvent analysis enables a direct assessment of the role of the nonlinear self-interactions of the screech mode.  Looking ahead, extending this framework to three-dimensional flows and diverse operating conditions could refine predictive models, paving the way for enhanced noise-control strategies in aerospace applications.  Beyond screeching jets, the workflow we exemplify in this paper could be used to understand the role of triadic interactions and nonlinear self-interactions in other flows containing a strong harmonic component.


\section*{Acknowledgements}
Simulations were performed on the University of Michigan's Great Lakes cluster using hours provided by the U-M Research Computing Package (UMRCP).

\section*{Declaration of Interests}
The authors report no conflict of interest.



\begin{appendices}

\section{Derivation of weight matrix based on Chu's energy norm} \label{appA}

The original Chu's weight matrix is derived as \citep{Hanifietal96, Schmidtetal18}
\begin{equation}
\bm W_c = \text{diag} \left(\frac{\bar T}{\gamma \bar \rho M_a^2}, \bar \rho, \bar \rho, \bar \rho, \frac{\bar \rho}{\gamma (\gamma - 1) \bar T M_a^2} \right),
\label{eqn:ChuTD}
\end{equation}
for the auxiliary variable set $\bm{q}_a$ consisting of density $\rho$, velocity components $u_a, v_a, w_a$, and temperature $T$. To obtain Chu's norm for $\bm q$ as defined in \eqref{eqn:statevariables}, the present derivation follows the approach of \citet{Bhattacharjeeetal24}, which builds upon the principles outlined by \citet{Karbanetal20}.

The nonlinear transfer function between our set of variables and the original set of variables is governed by
\begin{equation}
\left(\rho, u_a, v_a, w_a, T \right) = g\left(\xi, u, v, w, p \right) = \left(\frac{1}{\xi}, \frac{u}{M_a}, \frac{v}{M_a}, \frac{w}{M_a}, p \xi \gamma \right).
\label{eqn:transfer}
\end{equation}
From the definition, $\rho = \frac{1}{\xi}$ holds true. For the velocity components, $u = \frac{u_d}{a_{\infty}}$ and $u_a = \frac{u_d}{U_j}$, which implies $u = M_a u_a$. Here, $(\cdot)_d$ represents the dimensional form of the variable. Similarly, $v = M_a v_a$ and $w = M_a w_a$. The last term in \eqref{eqn:transfer} is derived from the ideal gas law,
\begin{equation}
p_d = \rho_d R T_d. 
\label{eqn:idealgas1}
\end{equation}
Substituting $p = \frac{p_d}{\bar \rho a_{\infty}^2}$, $\rho = \frac{\rho_d}{\bar \rho}$, and $T = \frac{T_d}{\bar T}$ into \eqref{eqn:idealgas1}, we obtain
\begin{equation}
p \bar \rho a_{\infty}^2 = \rho \bar \rho R T \bar T,
\label{eqn:idealgas2}
\end{equation}
which can be further simplified using $a_{\infty} = \sqrt{\gamma R \bar T}$ to
\begin{equation}
p = \frac{\rho T}{\gamma}, 
\label{eqn:idealgas3}
\end{equation}
where $\gamma = \frac{C_p}{C_v}$ is the ratio of specific heats, and $R$ is the ideal gas constant. Thus, $T = p \xi \gamma$ is proven.

The weight matrix is computed as
\begin{equation}
\bm W = \bm J^* \bm W_c \bm J,
\label{eqn:WeightDef}
\end{equation}
where
\begin{equation}
\bm J = \frac{\partial g}{\partial \bm q} \bigg|_{\bm q = \bar{\bm q}}
\label{eqn:Jacobian}
\end{equation}
is the Jacobian matrix, which is expressed as
\begin{equation}
\bm J = 
\begin{bmatrix}
-\frac{1}{\bar \xi} & 0 & 0 & 0 & 0 \\
0 & \frac{1}{M_a} & 0 & 0 & 0 \\
0 & 0 & \frac{1}{M_a} & 0 & 0 \\
0 & 0 & 0 & \frac{1}{M_a} & 0 \\
\gamma \bar p & 0 & 0 & 0 & \gamma \bar \xi
\end{bmatrix}.
\label{eqn:Jacobian_mat}
\end{equation}
The final weight matrix is then
\begin{equation}
\bm W = \frac{1}{M_a^2}
\begin{bmatrix}
\frac{\gamma \bar p}{(\gamma - 1) \bar \xi ^2} & 0 & 0 & 0 & \frac{1}{(\gamma - 1) \bar \xi} \\
0 & \frac{1}{\bar \xi} & 0 & 0 & 0 \\
0 & 0 & \frac{1}{\bar \xi} & 0 & 0 \\
0 & 0 & 0 & \frac{1}{\bar \xi} & 0 \\
\frac{1}{(\gamma - 1) \bar \xi} & 0 & 0 & 0 & \frac{1}{(\gamma - 1) \bar p}
\end{bmatrix}.
\label{eqn:weight2}
\end{equation}

\end{appendices}

\bibliographystyle{jfm}
\bibliography{Screeching}

\end{document}